\documentclass[pre,aps,twocolumn,amsmath,amssymb,floats,superscriptaddress,nofootinbib,10pt,showkeys]{revtex4-2}
\usepackage{graphicx}% Include figure files
\usepackage{dcolumn}% Align table columns on decimal point
\usepackage{color,xcolor}
\usepackage{bm,comment}% bold math
\usepackage[linktoc=none]{hyperref}
\newcommand*{\VEC}[1]{\boldsymbol{#1}}
\newcommand*{\TENSOR}[1]{\mathsf{#1}}
\newcommand*{\OP}[1]{{\cal{#1}}}

\usepackage[version=3]{mhchem}
\usepackage{units}

\begin{document}
\title{Single-molecule trajectories of reactants in chemically active biomolecular condensates
}
\author{Stefano Bo}\affiliation{Department of Physics, King's College
London, London WC2R 2LS, United Kingdom}\affiliation{Max Planck Institute for the Physics of Complex Systems, N\"othnitzer Stra\ss e 38, 01187 Dresden, Germany}
\author{Lars Hubatsch}\affiliation{Max Planck Institute for the Physics of Complex Systems, N\"othnitzer Stra\ss e 38, 01187 Dresden, Germany}
\affiliation{Max Planck Institute of Molecular Cell Biology and Genetics, Pfotenhauerstra{\ss}e 108, 01307 Dresden, Germany}
\author{Frank Jülicher}\affiliation{Max Planck Institute for the Physics of Complex Systems, N\"othnitzer Stra\ss e 38, 01187 Dresden, Germany}
\affiliation{Center for Systems Biology Dresden, Pfotenhauerstrasse 108, 01307 Dresden, Germany}
\affiliation{Cluster of Excellence Physics of Life, TU Dresden, 01062 Dresden, Germany}

%\date{April 2022}

\begin{abstract}
    Biomolecular condensates provide distinct chemical environments, which control various cellular processes. The diffusive dynamics and chemical kinetics inside phase-separated condensates can be studied experimentally by fluorescently labeling molecules, providing key insights into cell biology. We discuss how condensates govern the kinetics of chemical reactions and how this is reflected in the stochastic dynamics of labeled molecules. This allows us to reveal how the physics of phase separation influences the evolution of single-molecule trajectories and governs their statistics. We find that, out of equilibrium, the interactions that enable phase separation can induce directed motion and transport at the level of single molecules. Our work provides a theoretical framework to quantitatively analyze single-molecule trajectories in phase-separated systems.\end{abstract}
\maketitle
\section{Introduction}
The functions of living cells require the spatial organization of biochemical processes. Organelles provide distinct biochemical compartments and are typically bounded by a membrane. However, many membraneless compartments exist, which are dynamic assemblies of many proteins and nucleic acids and are called biomolecular condensates~\cite{Banani2017,Alberti2019}. These condensates dynamically exchange material with the surroundings and
 their composition and size are typically controlled. Disruptions in their organization have been linked to pathologies anddisease~\cite{Patel2015AMutation,Alberti2021BiomolecularAgeing,Tsang2020PhaseMutations}. 
 Many biological condensates can be understood as multi-component liquid-like droplets that coexist with the 
 surrounding medium through phase separation ~\cite{Hyman2014,Weber2019,Fritsch2021,Julicher2024DropletSeparation,Hubatsch2021QuantitativeCondensates}.
 Phase-separated droplets in a multicomponent mixture provide distinct environments that can govern the local chemistry~\cite{Bauermann2022b,Harmon2022MolecularDroplets,Nott2016MembranelessFilters,Shelest2024PhaseFluxes}.  
 Such droplets can serve as chemical microreactors and chemical processes can promote or inhibit condensate formation ~\cite{Hondele2019DEAD-boxOrganelles,Cotton2022Catalysis-InducedActivity,Zwicker2022TheSeparation,Dorner2024TheOrganelles}, enabling many feedback and control mechanisms~\cite{Zwicker2016GrowthProtocells,Weber2019,Kirschbaum2021ControllingReactions,Bauermann2022a,Sastre2025SizeCells}.

 Condensates in cells are often of a size smaller than 1$\mu\mathrm{m}$, which makes their study by conventional fluorescence microscopy challenging. Single-molecule experiments and super-resolution microscopy can provide a higher spatial and temporal resolution, which permits to reveal dynamic processes that are otherwise inaccessible. Probing systems at the single-molecule level can reveal aspects of cellular heterogeneity~\cite{Leake2013TheTime}.
 Furthermore, single-molecule experiments allow us to access fluctuations at small scales and study their propagation to larger scales. Motivated by this perspective, theoretical approaches were developed to understand the stochastic motion of single molecules in biomolecular condensates and across condensate boundaries~\cite{Bo2021a}, as well as their signatures at larger scales~\cite{Heltberg2021PhysicalSub-compartments}. 
 Recent experiments have observed single-molecule dynamics in condensates~\cite{Kent,Chappidi2024PARP1-DNAEnds,Erkamp2024DifferentialCondensates,Mine-Hattab2021}.So far, theoretical studies of single-molecule motion have not taken into account the effects of chemical reactions. Given the importance of condensates as
organizers of chemical processes (see {\it e.g.}~\cite{Nott2016MembranelessFilters,Bauermann2022a,Bauermann2022b,Chappidi2024PARP1-DNAEnds}), in the present work, we address the stochastic dynamics of single molecules that can undergo chemical reactions in a phase-separated system. 

% \mysection{Labeling molecules in a mixture}\label{sec:lab}
\section{Labeling molecules in a mixture}\label{sec:lab} 

 \begin{figure}
		\centering
			\includegraphics[width=\columnwidth]{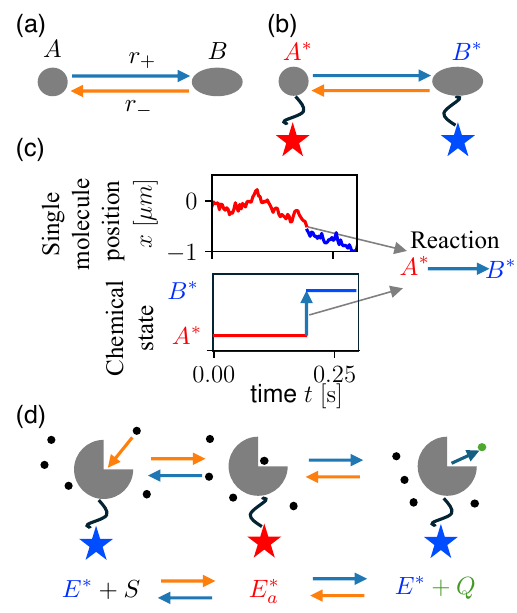}
					\caption{(a): A chemical reaction converts a molecule $A$ into $B$ with rate $r_+$ and back with rate $r_-$. (b): Labeling molecules $A$ and $B$ by attaching a fluorescent tag allows us to follow their dynamics and chemical kinetics and unveil their single-molecule properties. The label changes color when a chemical reaction occurs, allowing us to distinguish the labeled molecule $A^*$ from $B^*$. (c): Example of a stochastic trajectory of a single molecule, displaying a reaction converting $A$ into $B$, following Eq.~\eqref{eq:langevin}. (d): An enzymatic reaction converts a substrate $S$ into a product $Q$. A labeled enzyme catalyzes the reaction and switches between an active conformation ($E_a$) and an inactive one $E$.}
			\label{fig:intro}
\end{figure}

Single-molecule experiments are typically performed by labeling individual molecules, for instance, by attaching a fluorescent tag.
When labeled molecules are illuminated with laser light, the tags emit light. This light can be of different color depending on the 
chemical identity or conformation of the molecule (e.g., for Fluorescence Resonance Energy Transfer (FRET)) ~\cite{Lakowicz1999}.
 To illustrate how labeling can reveal behaviors of single molecules, let us consider a  mixture composed of 
 components $A$ and $B$ that can undergo a monomolecular chemical reaction in a solvent
%$
%A \underset{r_-}{\stackrel{r_{+}}{\rightleftharpoons}} B 
%$,
$A \rightleftharpoons B $,
see Fig.~\ref{fig:intro}a).
% \begin{align}\label{eq:rAB}
% A \underset{r_-}{\stackrel{r_{+}}{\rightleftharpoons}} B \, .
% \end{align}
%
We consider the case in which a labeled molecule of species $A$, denoted  $A^*$, emits light of one color, e.g. red.
When a chemical reaction  $A\to B$ occurs, the labeled molecule $A^*$ is converted into a labeled form of molecule $B$, denoted $B^*$.
If this new molecule $B^*$, emits light of a different color (e.g., blue), the two species are distinguishable,  see Fig.~\ref{fig:intro}b).
To develop a formalism that captures the stochastic dynamics in space of such labeled molecules undergoing chemical reactions, we first introduce the general framework and notation
used to describe multi-component phase separating mixtures.
 % \mysection{Chemical reactions in a phase-separating system}\label{sec:chem_ps}

 \section{Chemical reactions in a phase-separating system}\label{sec:chem_ps}
We consider an incompressible multicomponent fluid that we describe in terms of the volume fraction of the components $\phi_A$, $\phi_B$ (with solvent volume fraction $\phi_0=1-\phi_A-\phi_B$). For simplicity, we consider the species $A$ and $B$ to have the same molecular volume ($\nu=\nu_A=\nu_B$) so that the reaction conserves the volume.
The reaction flux, $r=r_+-r_-$,  %per unit volume 
where $r_+$ and $r_-$ denote the forward and backward contributions, is driven by differences in chemical potential and is governed 
by a detailed balance condition
\begin{align}\label{eq:detailed_balance}
\frac{r_{+}}{r_-} = e^{\frac{\mu_A-\mu_B}{k_B T}}  \,, 
\end{align}
where $k_B$ is the Boltzmann constant and the exchange chemical potentials are defined as functional derivatives of the free energy $F$ of the system as $\mu_\alpha=\nu \delta F/\delta \phi_\alpha$, for $\alpha=A,\,B$ (see Eq.~\eqref{eq:chem_potAB}). We express the free energy as
\begin{align}\label{eq:free_energyAB}
    &F[\phi_A,\phi_B]=\\\nonumber
    &\int d^\text{d} x \, \left(  f(\phi_A,\phi_B)  +  \frac{\kappa_A}{2\nu}\left(\nabla\phi_A \right)^2 +\frac{\kappa_B}{2\nu}\left(\nabla\phi_B \right)^2 \right) \, ,
\end{align}
where $\kappa_A$, $\kappa_B$ characterize the contributions of concentration 
gradients to the free energy, which play a role in setting the interface width and govern surface tensions. 
The free energy density of a homogenous phase can be written as $f=u-Ts$, where the density of the entropy of mixing is
\begin{align}\label{eq:sAB}
s =-\frac{k_B}{\nu}(\phi_A\ln\phi_A  + 
\phi_B\ln\phi_B+
\frac{\nu}{\nu_0}\phi_0\ln\phi_0)\,,
\end{align}
with $\phi_0=1-\phi_A-\phi_B$ and  with $\nu_0$ being the molecular volume of the solvent.
The internal free energy density $u$ accounts for the interaction of the components, described using a Flory-Huggins model by coefficients $\chi_{\alpha\beta}$  
(see Eq.~\eqref{eq:free_energy_compactAB}~)\cite{Krapivsky2010,DeGennes1979ScalingPress.}. 
These interactions balance the entropy of mixing and can therefore lead to phase separation. %shown in Eqs.~\eqref{eq:free_energyAB}, \eqref{eq:free_energy_compactAB}.
We express the reaction fluxes as
\begin{align}\label{eq:reaction_ratesAB}
    r_+=ke^{\frac{\mu_A}{k_BT}}\,,\qquad \qquad r_-=ke^{\frac{\mu_B}{k_BT}}\,,
\end{align}
where the rate $k$ is a reaction rate, %per unit volume, 
which, in general, depends on composition.
The dynamics of the volume fractions obey the balance equations
\begin{align}  \label{eq:cont_AB}
   \partial_t \phi_A = - \nabla \cdot \boldsymbol{j}_A-\left(r_+-r_-\right) \\\nonumber
   \partial_t \phi_B = - \nabla \cdot \boldsymbol{j}_B+\left(r_+-r_-\right)
\end{align}
where, following linear irreversible thermodynamics,
% \begin{align}
%     r=r^+-r_-
% \end{align}
%  is the source rate due to chemical reactions, and 
 the diffusive fluxes are driven by gradients in the exchange chemical potentials  and governed by a composition-dependent mobility matrix $M$, 
%  \begin{align}\label{eq:j}
% \boldsymbol{j}_\alpha =-  \nu\sum_\beta M_{\alpha\beta}\nabla\mu_\beta   \,,
% \end{align}
 \begin{align}\label{eq:j}
\boldsymbol{j}_\alpha =-  \sum_\beta M_{\alpha\beta}\nabla\mu_\beta   \,,
\end{align}
 where the sum runs over $\beta=A,\, B$. Here and in the following, we denote vectors by bold symbols.
 We consider a simple mobility matrix $M$  that gives rise to a diagonal diffusion matrix in the dilute limit\cite{Kramer1984,Mao2020,Bo2021a,Cotton2022}
 % $M_{AA}= \phi_A(m_{A0}\phi_0 +
 % m_{AB} \phi_B)$, $M_{BB}= \phi_B(m_{B0}\phi_0 +
 % m_{AB} \phi_A)$, $M_{AB}=M_{BA}=-m_{AB}\phi_A \phi_B$
 \begin{align}\label{eq:MAB}
     M=\left( \begin{array}{c c }
\phi_A(m_{A0}\phi_0 +
 m_{AB} \phi_B)& -m_{AB}\phi_A \phi_B	 \\
-m_{AB}\phi_A \phi_B & \phi_B(m_{B0}\phi_0 +
 m_{AB} \phi_A)
\end{array}
\right)\,.
 \end{align}
 The coefficients $m_{\alpha\beta}=m_{\beta\alpha}$ describe how rapidly molecules of species $\alpha$ and $\beta$  exchange position; $m_{A0}$ and $m_{B0}$ are the mobility coefficients of species $A$ and $B$ in the solvent, respectively. Here, $m_{\alpha\beta}$ is
a $3\times3$ symmetric matrix containing 6 independent coefficients. Only 3 of these coefficients are relevant to describe the evolution of the volume fraction of species $A$ and $B$ and appear in the mobility matrix $M$ in Eq.~\eqref{eq:MAB}. The other coefficients ($m_{AA}$, $m_{BB}$ and $m_{00}$) describe how molecules of the same species exchange position. These exchanges do not modify the volume fraction profiles and the coefficients are therefore not featured in the mobility matrix $M$. However, as we shall see in the following, they play a key role in the dynamics of labeled molecules and single molecules. 
The explicit expressions of the fluxes defined in Eq.~\eqref{eq:j} with the mobility of Eq.~\eqref{eq:MAB} are given in Eq.~\eqref{eq:jAB_app}.
This general formalism takes into account the interaction energies of the components and the chemical reactions among the components in a thermodynamically consistent way and allows us to describe chemically active biomolecular condensates near equilibrium~\cite{Bauermann2022a,Bauermann2022b}.

 \section{Dynamics of labeled molecules in a reacting mixture}
 The dynamics of the labeled molecules (and of the unlabeled ones as well) can be described within the general formalism that we outlined above. %in Section \ref{sec:chem_ps}.
 We denote the labeled (unlabeled) volume fractions by $\phi_A^*$ and $\phi_B^*$ ($\overline{\phi}_A$  and $\overline{\phi}_B$).  The total volume fraction of species $A$ is $\phi_{A}=\phi^*_{A}+\overline{\phi}_{A}$ and for $B$, $\phi_{B}=\phi_{B}^*+\overline{\phi}_{B}$.
 The reaction fluxes for labeled (unlabeled) molecules are $r_+^*$ and $r_-^*$  ($\overline{r}_+$ and $\overline{r}_-$).
Considering, for simplicity, that labeling does not change the physical properties of molecules such as the interaction parameters $\chi$, the reaction rate $k$, and $\kappa_\alpha$, we derive the dynamics of labeled molecules. 
 Labeling allows us to distinguish molecules and affects the mixing entropy  \cite{VanKampen1984}, which now reads
\begin{align}\label{eq:sAB_labeled}\nonumber
s =-\frac{k_B}{\nu}\big(&\phi_A^*\ln\phi_A^*  +\overline{\phi}_A\ln\overline{\phi}_A+
\\
&
\phi_B^*\ln\phi_B^*+
\overline{\phi}_B\ln\overline{\phi}_B+
\frac{\nu}{\nu_0}\phi_0\ln\phi_0\big)\,,
\end{align}
where it is understood that $\phi_0=1-\phi_A^*-\overline{\phi}_A-\phi_B^*-\overline{\phi}_B$.
The resulting exchange chemical potentials for labeled and unlabeled molecules are
 % \begin{align}\label{eq:chemical_potential_AB_lab}
 %     \mu_A^*=\mu_A+k_BT\ln\frac{\phi_A^*}{\phi_A}\,,\;\; 
 %     \mu_B^*=\mu_B+k_BT\ln\frac{\phi_B^*}{\phi_B}\,,\\\nonumber
 %      \overline{\mu}_A=\mu_A+k_BT\ln\frac{\overline{\phi}_A}{\phi_A}\,,\;\;
 %     \overline{\mu_B}=\mu_B+k_BT\ln\frac{\overline{\phi}_B}{\phi_B}\,,
 % \end{align}
  \begin{align}\label{eq:chemical_potential_AB_lab}
     \mu_\alpha^*=\mu_\alpha+k_BT\ln\frac{\phi_\alpha^*}{\phi_\alpha}\,,\;\; 
      \overline{\mu}_\alpha=\mu_\alpha+k_BT\ln\frac{\overline{\phi}_\alpha}{\phi_\alpha}\,,
 \end{align}
 where $\mu_\alpha$ is the exchange chemical potential for species $\alpha$  if labeled and unlabeled molecules are not distinguished, given in Eq.~(\ref{eq:chem_potAB}).
 
 \subsection{Reactions in a labeled ternary mixture}
From the definition of the reaction fluxes~\eqref{eq:reaction_ratesAB}, and the expressions for the exchange chemical potential for labeled molecules~\eqref{eq:chemical_potential_AB_lab},we obtain the
  the reaction fluxes for the labeled and unlabeled molecules 
 \begin{equation}\label{eq:rr*}
     r_+^*=r_+\frac{\phi_A^*}{\phi_A},\; r_-^*=r_-\frac{\phi_B^*}{\phi_B},\;
      \overline{r}_+=r_+\frac{\overline{\phi}_A}{\phi_A} ,\;  \overline{r}_-=r_-\frac{\overline{\phi}_B}{\phi_B}
     \,,
 \end{equation}
  where we have used that labeling does not change the reaction rates $k$.
 This relation shows that the sum of labeled and unlabeled fluxes equals the reaction fluxes for the case when labeled and unlabeled molecules are not distinguished $r_+=r_+^*+\overline{r}_+$ and $r_-=r_-^*+\overline{r}_-$. This confirms that, if the labeling does not change the physical parameters of the systems, the overall fluxes are also unchanged.
 
\subsection{Diffusion flux in a ternary mixture containing labeled molecules}
The diffusive flux of labeled molecules %, we make use of linear irreversible thermodynamics and compute the flux 
follows from Eq.~\eqref{eq:j} with the exchange chemical potentials for labeled and unlabeled molecules of Eqs.~\eqref{eq:chemical_potential_AB_lab}, where the sum now runs over the four components $A$, $A^*$, $B$, and $B^*$.
%
%$\boldsymbol{j}_\alpha =-  \nu\sum_\beta M_{\alpha\beta}\nabla\mu_\beta $ from . %T now we have four components and the solvent.
Compared to the case without labeling, we now need additional information regarding how quickly molecules of species $A$ exchange positions with molecules of the same kind, {\it i.e.}, the coefficients $m_{AA}$
and $m_{BB}$. 
Indeed, the mobility matrix for this labeled ternary mixture is now a $4\times 4$ matrix [see Eq.~\eqref{eq:MAB_labeled}] featuring 5 kinetic coefficients with $m_{AA}$, $m_{BB}$ in addition to $m_{A0}$, $m_{AB}$ and $m_{B0}$.
 The flux of labeled molecules of species $\alpha= A,\, B$  %$A$ %($B)$ 
 reads
 \begin{subequations}
\begin{align}\label{eq:j_multiAB}
    \boldsymbol{j}_{\alpha}^* =& D_\alpha\left(-\nabla \phi_\alpha^*+ \phi_\alpha^*\frac{\nabla\phi_\alpha}{\phi_\alpha}\right)+\frac{\phi_\alpha^*}{\phi_\alpha}\boldsymbol{j}_\alpha
    % \boldsymbol{j}_{A}^* =& D_A\left(-\nabla \phi_A^*+ \phi_A^*\frac{\nabla\phi_A}{\phi_A}\right)+\frac{\phi_A^*}{\phi_A}\boldsymbol{j}_A
    %\\\nonumber
    %\boldsymbol{j}_{B}^* =& D_B\left(-\nabla \phi_B^*+ \phi_B^*\frac{\nabla\phi_B}{\phi_B}\right)+\frac{\phi_B^*}{\phi_B}\boldsymbol{j}_B\,,
\end{align}
where $\boldsymbol{j}_\alpha$
 %($\boldsymbol{j}_B$) 
is the diffusive flux for molecule $\alpha$ %($B$) 
when labeled and unlabeled molecules are not distinguished given in Eq.~\eqref{eq:j}, which vanishes at equilibrium. This expression coincides with the one found for the case without chemical reactions in Refs.~\cite{Hubatsch2021QuantitativeCondensates,Bo2021a}. %and is shown in Fig.\ref{fig:neq}b.
Analogously, the flux of unlabeled molecules
%
% At equilibrium, these dynamics correspond to diffusion under the influence of an effective potential %$W_\alpha=-k_BT\ln\phi_\alpha $, 
% \begin{align}\label{eq:W}
%   W_\alpha=-k_BT\ln\phi_\alpha \,,
% \end{align}
% which takes into account the overall effect of the intermolecular forces, which can lead to phase separation.
%$W_\alpha=-k_BT\ln\phi_\alpha$. 
reads 
\begin{align}%\label{eq:j_multiAB}
    \overline{\boldsymbol{j}}_{\alpha} =& D_\alpha\left(-\nabla \overline{\phi_\alpha}+ \overline{\phi_\alpha}\frac{\nabla\phi_\alpha}{\phi_\alpha}\right)+\frac{\overline{\phi_\alpha}}{\phi_\alpha}\boldsymbol{j}_\alpha\,,
    % \boldsymbol{j}_{A}^* =& D_A\left(-\nabla \phi_A^*+ \phi_A^*\frac{\nabla\phi_A}{\phi_A}\right)+\frac{\phi_A^*}{\phi_A}\boldsymbol{j}_A
    %\\\nonumber
    %\boldsymbol{j}_{B}^* =& D_B\left(-\nabla \phi_B^*+ \phi_B^*\frac{\nabla\phi_B}{\phi_B}\right)+\frac{\phi_B^*}{\phi_B}\boldsymbol{j}_B\,,
\end{align}
\end{subequations}
confirming that the total flux of molecules $\alpha$, (summing labeled and unlabeled ones) equals the flux of molecules $\alpha$, when labeled and unlabeled molecules are not distinguished $\boldsymbol{j}_{\alpha}=\boldsymbol{j}_\alpha^*+\overline{\boldsymbol{j}}_\alpha$. % with $\boldsymbol{j}_\alpha$ given in Eq.~\eqref{eq:j}.
%$\boldsymbol{j}_{A}=\boldsymbol{j}_A^*+\overline{\boldsymbol{j}}_A$. %($\boldsymbol{j}_{B}=\boldsymbol{j}_B^*+\overline{\boldsymbol{j}}_B$).
The single-molecule diffusion coefficients are %$D_A=k_BTm_A$, %g_Am_{A0}$ 
%($D_B=k_BTm_{B}$)
% \begin{align}\label{eq:d}
% D_A=&k_BT
% g\;m_{A0}\,,
% \end{align}
%where $m_A$ %($m_B$) 
%is the single-molecule mobility, which, in general, depends on the composition as
 % and the dimensionless coefficient $g_A$ ($g_B$) describes interactions with other components. In general, $g$ depends on composition and is given by
 \begin{subequations}\label{eq:DAB}
 \begin{align}\label{eq:DA}
D_A(x,t)=&k_BT\left(m_{A 0}\phi_0+m_{A A}\phi_A+m_{A B}\phi_B\right)\,,
\\\label{eq:DB}
D_B(x,t)=&k_BT\left(m_{B 0}\phi_0+m_{A B}\phi_A+m_{B B}\phi_B\right)
%\,,
%g_B=&1-\xi_{B B}\phi_B-\xi_{BA}\phi_A
\end{align}
 \end{subequations}
and, in general, depend on the composition and are different from the collective diffusion coefficients, see Eq.~\eqref{eq:D_col}.
Notice that these diffusion coefficients require additional kinetic parameters ($m_{AA}$
and $m_{BB}$) than the three appearing in the mobility matrix~\eqref{eq:MAB}, which were sufficient to describe the motion of molecules without labeling.
% \begin{align}\label{eq:gAB}
% g_A=&1-\left(1-\frac{m_{A A}}{m_{A 0}}\right)\phi_A-\left(1-\frac{m_{A B}}{m_{A 0}}\right)\phi_B\\
% \nonumber
% g_B=&1-\left(1-\frac{m_{A B}}{m_{B 0}}\right)\phi_A-\left(1-\frac{m_{B B}}{m_{B 0}}\right)\phi_B\,,
% %g_B=&1-\xi_{B B}\phi_B-\xi_{BA}\phi_A
% \end{align}
% \begin{align}\label{eq:mgAB}
% m_{A0}g_A=&m_{A0}-(m_{A0}-m_{AA})\phi_A-(m_{A0}-m_{AB})\phi_B\\
% \nonumber
% g_B=&1-\xi_{B B}\phi_B-\xi_{BA}\phi_A
% \end{align}

\subsection{Reaction-diffusion dynamics of labeled molecules in a ternary mixture}
Combining Eq.\eqref{eq:rr*} and \eqref{eq:j_multiAB}, we obtain the equations governing the reaction-diffusion dynamics for the labeled molecules of the two species
 \begin{subequations}\label{eq:effective_rate_AB}
\begin{align}\label{eq:effective_rate_A}
\partial_\mathrm{t}\phi_A^*
=&%k_\mathrm{B}T \,
\nabla\cdot\left[D_{A}\left(\nabla \phi_A^*-\phi_A^*\frac{\nabla\phi_A}{\phi_A}\right)-\phi_A^*\frac{\boldsymbol{j}_A}{\phi_A}\right]\\\nonumber
&-r_+\frac{\phi_A^*}{\phi_A}+r_-\frac{\phi_B^*}{\phi_B}\, ,\\\label{eq:effective_rate_B}
% \phi_A^*\frac{\boldsymbol{j}_A}{\phi_A}\right]-\frac{r^_{A\to B}}{\phi_A}\phi_A^*+\frac{r^-_{A\to B}}{\phi_B}\phi_B^*\, ,
\partial_\mathrm{t}\phi_B^*
=&%k_{B}T \,
\nabla\cdot\left[D_{B}\left(\nabla \phi_B^*-\phi_B^*\frac{\nabla\phi_B}{\phi_B}\right)-\phi_B^*\frac{\boldsymbol{j}_B}{\phi_B}\right]\\\nonumber
&-r_-\frac{\phi_B^*}{\phi_B}+r_+\frac{\phi_A^*}{\phi_A}\, ,
% \phi_B^*\frac{\boldsymbol{j}_B}{\phi_B}\right]-\frac{r^-_{A\to B}}{\phi_B}\phi_B^*+\frac{r^+_{A\to B}}{\phi_A}\phi_A^*\, .
%\label{eq:effective_rate_B}
\end{align}
\end{subequations}
where the dynamics of $\phi_A$ and $\phi_B$ is described by \eqref{eq:cont_AB}.
These expressions are the first main results of the paper.

 \section{Stochastic description of single molecules}
  Equations~\eqref{eq:effective_rate_AB} describe the reaction-diffusion dynamics of the labeled molecules. Taking the limit where only a small number of molecules are labeled, we can interpret Eqs.~\eqref{eq:effective_rate_AB} as describing the evolution of the probability density of having a single labeled molecule of species $A$ ($B$) in position $x$ at time $t$: $P_\alpha(x,t)=\phi_\alpha^*/\int \phi_\alpha^*dx$ where $\alpha$ is either $A$ or $B$.
% \begin{align}
%     P_A(x,t)=\frac{\phi_A^*}{\int \phi_A^*dx}\qquad P_B(x,t)=\frac{\phi_B^*}{\int \phi_B^*dx}\,.
% \end{align}
The probabilities $P_A(x,t)$ and $P_B(x,t)$ obey the Fokker-Planck equations 
\begin{subequations}\label{eq:FPAB}
\begin{align}\label{eq:FPA}
\partial_t P_A &= -\nabla\cdot\left( \VEC{v}_A P_A- D_A\nabla P_A\right) +K_-P_B-K_+P_A \\\label{eq:FPB}
\partial_t P_B &= -\nabla\cdot \left(\VEC{v}_B P_B- D_B\nabla P_B\right)+K_+P_A-K_-P_B\,.
\end{align} 
\end{subequations}
Here, $\VEC{v}_A$ and $\VEC{v}_B$ are the single-molecule deterministic drift velocities if the tracked molecule has the chemical identity $\alpha=A,\,B$,
\begin{align}\label{eq:vAB}
    \VEC{v}_\alpha(x,t)=&-\frac{D_\alpha}{k_BT} \nabla W_\alpha + \frac{\VEC{j}_\alpha}{\phi_\alpha}\,,
%    \\\nonumber
%    \VEC{v}_B=&-\frac{D_B}{k_BT} \nabla W_B + \frac{\VEC{j}_B}{\phi_B}
\end{align}
where the first term is driven by the gradient of the corresponding effective potential $ W_\alpha(x,t)=-k_BT\ln\phi_\alpha$,
% \begin{align}\label{eq:W}
%   W_\alpha(x,t)=-k_BT\ln\phi_\alpha \,,
% \end{align}
which takes into account the overall effect of the intermolecular forces that can lead to phase separation~\cite{Bo2021a}.
The second term in Eq.~(\ref{eq:vAB}) is due to the non-equilibrium volume-fraction flux $\VEC{j}_\alpha$ defined in Eq.~(\ref{eq:j}).
$D_\alpha$ is the single-molecule diffusion coefficient of species $\alpha$, given in Eq.~(\ref{eq:DAB}), which can depend on composition.
Finally,
\begin{align}\label{eq:K}
    K_+(x,t)=\frac{r_+}{\phi_A}\qquad \qquad  K_-(x,t)=\frac{r_-}{\phi_B}
\end{align}
are the single-molecule reaction rates transforming species $A$ into $B$ and $B$ into $A$, respectively (see Ref. \cite{gillespie1992rigorous}). These rates depend on composition as they are influenced by intermolecular interactions.
Importantly, and at variance with the dilute case,  the ratio of the rates $K^{eq}_+/K^{eq}_-=\phi_B^{eq}/\phi_A^{eq}$ satisfies the equilibrium condition only at equilibrium concentrations. %, which is, in that case, independent of composition.
The stochastic process described by the Fokker-Planck equation~(\ref{eq:FPAB}) can also be described by the coupled (Ito) Langevin equations for molecular trajectories
\begin{subequations}\label{eq:langevin}
\begin{align}\label{eq:langevin_x}
    d\VEC{x}&=\VEC{v}(\sigma,\,\VEC{x})dt+\nabla D(\sigma,\,\VEC{x})dt+\sqrt{2D(\sigma,\VEC{x})}\;d\VEC{w}\,\\\label{eq:langevin_sigma}
    d\sigma&=d\eta_+(\sigma,x)-d\eta_-(\sigma,x)
\end{align}
\end{subequations}
where $\VEC{x}(t)$ is the stochastic time-dependent position of a tracked molecule. The stochastic variable $\sigma(t)$, which only takes discrete values $\sigma=0,\,1$, describes the instantaneous chemical state of the molecule. Here $0$ and $1$ represent chemical species $A$ and $B$, respectively.
The drift velocity  $\VEC{v}(\sigma_t,x_t)=\VEC{v}_A\delta_{\sigma_t,0}+\VEC{v}_B\delta_{\sigma_t,1}$ and the diffusion coefficient $D(\sigma_t,x_t)=D_A\delta_{\sigma_t,0}+D_B\delta_{\sigma_t,1}$ depend on position $x$ and on the chemical state $\sigma$.
Molecular noise is described by the stochastic variables $\VEC{w}$, $\eta_+$ and $\eta_-$.  The Gaussian white noise $\VEC{w}$ vanishes on average ($ \langle d\VEC{w}_t\rangle=0$), its fluctuations during a time increment $dt$  
satisfy $\langle dw^i dw^j\rangle=dt\delta_{ij}$ %$\langle\xi_\alpha(t)\xi_\beta(t')\rangle=\delta(t-t')\delta_{\alpha\beta}$ ().
% \begin{align}
%     \langle dw^\alpha dw^\beta\rangle=dt\delta_{\alpha\beta}\,,
% \end{align}
and $d\VEC{w}$ at different times are uncorrelated. The indices
$i,\,j$ correspond to the spatial coordinates \cite{Gardiner1985}.
The chemical noise is a Poisson process that only takes increments $d\eta^\pm=0,\,1$ with average $ \langle d\eta^+\rangle=K_+\delta_{\sigma_t,0}dt$, $\langle d\eta^-\rangle=K_-\delta_{\sigma_t,1}dt$
% \begin{align}
%     \langle d\eta^+\rangle=K_+\delta_{\sigma_t,0}dt\qquad \langle d\eta^-\rangle=K_-\delta_{\sigma_t,1}dt
% \end{align}
during a time increment $dt$, set by the rates rates $K_+$ and $K_-$.
% It describes the increments of the point process obeyed by the chemical reactions, which occur with rates $K_+$ and $K_-$.
The coupled Langevin equations describe the particle dynamics as it diffuses, subject to the deterministic drift and as it switches between chemical states $A$ and $B$. An example of these trajectories is shown in Fig.~\ref{fig:intro}c.
 The multiplicative noise is taken in the Ito discretization and $\nabla D$ is the associated noise-induced drift.
\section{Single molecules in coexisting equilibrium phases}
\begin{figure}[h!]
		\centering
            \includegraphics
            [trim= 0.1cm 0cm 0cm 0cm, clip=true, width=\columnwidth]
            {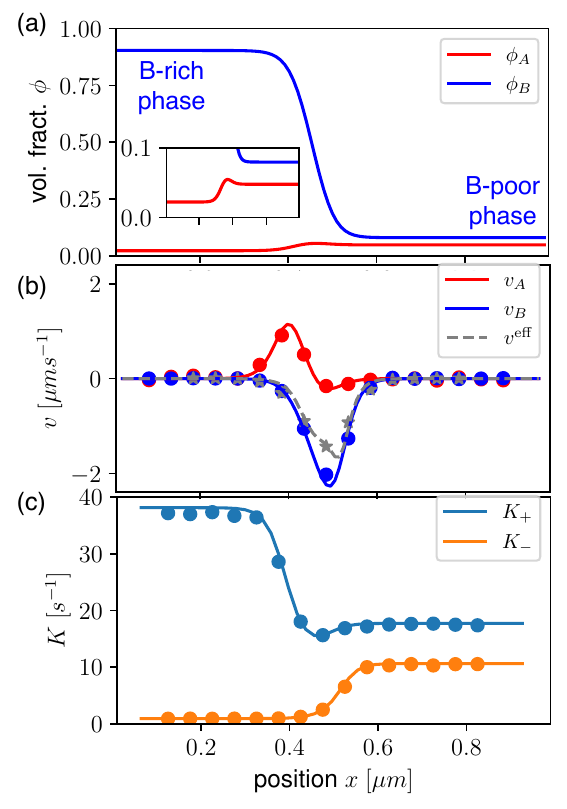}
						\caption{
                        \textbf{Single molecules in an equilibrium reacting ternary mixture.}
      (a): Volume fraction profiles for two components $A$ and $B$ in an equilibrium system. For the chosen parameters (see Appendix \ref{app:num_eq}), $B$ phase separates into a $B$-poor phase (low volume fraction) and a $B$-rich one (high volume fraction). $A$ is more dilute and is enriched in the $B$-poor phase.
      (b):
       Single-molecule drift profiles for molecule $A$ and $B$ as given in Eq.~(\ref{eq:drifts}). 
      The solid symbols are mean local velocities for chemical species ($A$ and $B)$, as defined in Eq.~\ref{eq:mean}, 
      for $\Delta t=\unit[0.001]{s}$. The gray dashed line is the effective drift following Eq.~\eqref{eq:veff} and the gray stars are local mean velocities (irrespective of molecular identity), as defined in Eq.~\ref{eq:mean_irr}, for the longer time step $\Delta t=\unit[0.01]{s}$.
 (c): Single-molecule reaction rate profiles following Eq.~(\ref{eq:K}). The solid symbols are measured from the numerical simulations of Eq~\eqref{eq:langevin} with a $\Delta t=\unit[0.001]{s}$.}  
			\label{fig:eq}
\end{figure}
As a first example, we consider a one-dimensional equilibrium system with two coexisting phases as shown by the volume fraction profiles in Fig.~\ref{fig:eq}(a). 
On the left side of the domain, there is a phase, which is rich in component $B$ and poor in component $A$, separated from a phase poor in $B$ and enriched in $A$ on the right.
Component $A$ is less abundant than component $B$. Both components display volume fraction gradients at the interface between the two phases, which, due to the molecular interaction, induce a deterministic drift on the molecules of both species, according to Eq.~\eqref{eq:vAB}, where $j_A=j_B=0$ because the system is in equilibrium. 
To illustrate our findings, we consider the case where components $A$ and $B$ have the same molecular volume as the solvent $\nu=\nu_0$, and have a constant single-molecule diffusion coefficient $D=mk_BT$ (independent of composition and position).
Since $j_\alpha=0$, the drift from Eq.~\eqref{eq:vAB} simplifies to $v_\alpha=D\nabla\phi_\alpha/\phi_\alpha$.
% The general expression of the drifts is given in Eq.~\eqref{eq:drifts} and the one for this example in Eq.~\ref{eq:drifts_eq}.
Since the two species are enriched in the opposite phases, the gradients have opposite signs, and this induces deterministic drifts in opposite directions: molecules of species $A$ experience a drift towards the right, while $B$ molecules towards the left, as shown in Fig.~\ref{fig:eq}(b).
These drifts can be measured by computing the mean displacements covered by single molecules as a function of their position, as detailed in Eq.~\eqref{eq:mean}. We report such measurements for simulated trajectories as solid circles in Fig.~\ref{fig:eq}(b).
Since the system is at equilibrium, the reaction rates from Eq.\eqref{eq:reaction_ratesAB} perfectly balance. Despite the difference in volume fraction between phases, the reaction rates are the same in both phases, in accordance with Ref.~\cite{Bauermann2022b}. As shown in Fig.~\ref{fig:eq}(c), the single-molecule reaction rates are different in the two phases following Eq.\eqref{eq:K}. 
These rates can be measured by counting the mean number of observed single-molecule reactions as a function of position. We show these measurements for simulated trajectories as solid circles in Fig.~\ref{fig:eq}(c).
\section{Single-molecule dynamics of a labeled enzyme}
As a second example, we discuss the dynamics of an enzyme catalyzing a reaction converting a substrate $S$ into a product $Q$:
% \begin{align}
%     \ce{E + S <=>T[$k_1$][$k_{-1}$] E_a  ->T[$k_2$] E + P}
% \end{align}
% \begin{align}
% E + S \underset{k_{-1}}{\stackrel{k_1}{\rightleftharpoons}}E_a  \underset{k_{-2}}{\stackrel{k_2}{\rightleftharpoons}}E + P \, ,
% \end{align}
\begin{align}
E + S {\rightleftharpoons}E_a  \rightleftharpoons E + Q \, ,
\end{align}
%$E + S {\rightleftharpoons}E_a  \rightleftharpoons E + Q$, 
as shown in Fig.~\ref{fig:intro}d).
In this process the enzyme switches between the active ($E_a$) and inactive ($E$) conformation.
A full description of the system would involve four components and the solvent.
To streamline the discussion, we do not describe the dynamics of the substrate %$S$ 
and product explicitly but follow the trajectories of the enzyme. %$Q$ 
%and model their effective influence by a position-dependent reaction rate $k(x)$.
This simplification leads us to a system featuring a ternary mixture. Identifying the active conformation $E_a$ with species $A$ and the inactive one $E$ with species $B$, we can describe the dynamics with the formalism we introduced above.
The fluxes rates $r_+$ and $r_-$ depend on the local amount of $A$ and $B$\footnote{The influence of the profile of substrate and product can be modeled by a position-dependent reaction rate $k(x)$, or reference chemical potential $\omega_\alpha(x)$ but in the following, we focus on constant $k$ and $\omega_\alpha$.}. %(and $\omega_\alpha(x)$?? we assumed it constant in the general derivations.t).

\subsection{Enzymes in concentration gradients:\\ directed motion}
As for the previous example, we assume $\nu=\nu_0$, a constant single-molecule diffusion coefficient, 
and study the system numerically. We start by analyzing the drift velocities defined in Eq.~\eqref{eq:vAB}, which are the deterministic components of the single-molecule dynamics.
The explicit expression can be written as 
\begin{align}\label{eq:v_alpha}
    v_\alpha=m\left[\nu\tilde{\chi}\sum_\beta g_{\alpha\beta}\nabla\phi_\beta+\tilde{\kappa}\sum_\beta c_{\alpha\beta}\nabla\nabla^2\phi_\beta\right]
\end{align}
where $\tilde{\chi}$ denotes the magnitude of the interactions as defined in Eq.~\eqref{eq:chitilde} and $\tilde{\kappa}=(\kappa_A+\kappa_B)/2$ is the mean gradient penalty. $g_{\alpha\beta}$ and $c_{\alpha\beta}$ depend on composition and are given in Eq.~\eqref{eq:g}, Eq.~\eqref{eq:c}, 
respectively.
%and the contribution $\mathcal{C}_\alpha$ depends on higher derivatives of the volume fraction profile and is given in Eq.~(\ref{eq:curva}). 
These drifts are driven by gradients of the volume fraction profiles and depend on the intermolecular interactions $\tilde{\chi}$ and gradient penalties $\tilde{\kappa}$.
The first important observation is that an enzyme in a concentration (or volume fraction) gradient of enzymes will experience a drift velocity if it interacts with other enzymes (\textit{i.e.} $\chi\neq0$).
In the absence of interactions ($\tilde{\chi}=0$, $\tilde{\kappa}=0$), the concentration gradients do not induce a drift, highlighting how the single-molecule drift velocity is distinct from diffusion fluxes for concentration fields. 
The relative magnitude of these drifts compared to diffusive effects is captured by the Peclet number Pe$_\alpha\equiv Lv_\alpha/D_\alpha$, where L is the size of the system,  %For molecule $A$ 
%In a system of size $L$, the Peclet number is
\begin{align}
    \label{eq:peclet}
    \text{Pe}_\alpha=%&\equiv L\frac{v_\alpha}{D_\alpha}\\\nonumber
    %&= 
    L \left[\frac{\nu\tilde{\chi}}{k_BT}\sum_\beta g_{\alpha\beta}\nabla\phi_\beta+
    \frac{\tilde{\kappa}}{k_BT}\sum_\beta c_{\alpha\beta}\nabla\nabla^2\phi_\beta
    \right]\,.
\end{align}
% \begin{align}
%     \label{eq:peclet}
%     Pe&\equiv L\frac{v_A}{D_A}\\\nonumber
%     &= L \frac{\nu\chi}{k_BT}\left[\phi_B\nabla\phi_A -\left(1-\phi_A\right)\nabla\phi_B\right] +L\frac{\mathcal{C}_A}{mk_BT}\,,
% \end{align}
The Peclet number in this system is governed by the ratio between the interaction and the thermal energy (and the gradient contributions to the free energy $\tilde{\kappa}$).\\
The second important observation is that the enzyme in the active conformation experiences a different drift from the inactive one. When $D_A=D_B$ and set equal to $D=mk_BT$, 
plugging the expression for the fluxes, Eq.~\eqref{eq:jAB_app}, into the definition of the drift~\eqref{eq:vAB} shows that the difference in the drifts reads
\begin{align}\label{eq:vA-vB_mu}
    v_A-v_B=m\nabla\left(k_BT\ln{\frac{\phi_A}{\phi_B}}-\mu_A+\mu_B\right)\,.
\end{align}
Interestingly, combining the local detailed balance condition, Eq.~\eqref{eq:detailed_balance}, with the definition of  the single-molecule reaction rates, Eq.~\eqref{eq:K}, reveals that  the difference in the drifts is related to the gradients of the reaction rates
\begin{align}\label{eq:vandK}
    v_A-v_B=D\nabla \ln\frac{K_-}{K_+}\,.
\end{align}

\subsection{Long-term effective drift and diffusion}
\begin{figure}[t]
		\centering
 			\includegraphics[trim= 0.3cm 0cm 0cm 0cm, clip=true, width=\columnwidth]{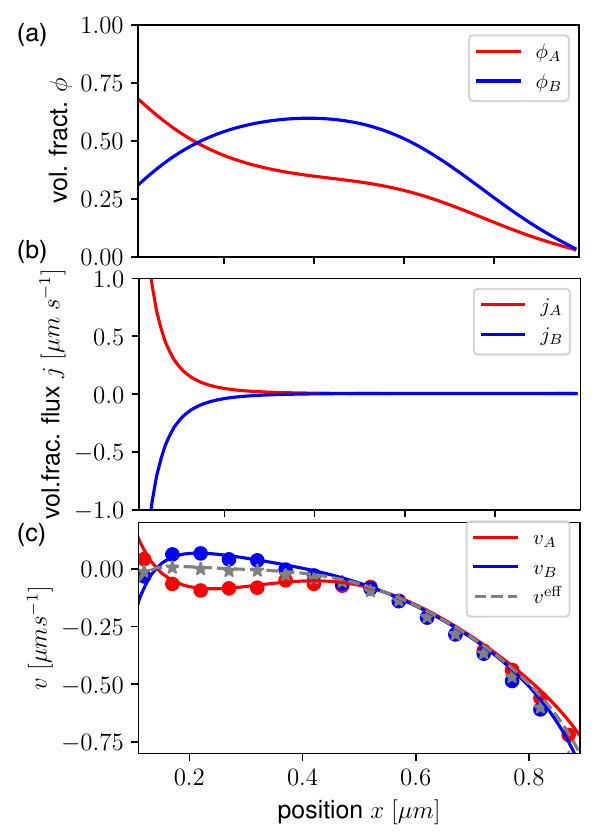}
					\caption{%{\textbf{Nonequilibrium volume fraction dynamics.} 
    \textbf{Nonequilibrium dynamics in a reacting ternary mixture.}
     (a): Volume fraction profiles for two components $A$ and $B$. (b): the solid lines are the volume fraction flux profiles $j_A$ and $j_B$ obtained via Eq.~(\ref{eq:j}). %The dashed lines are reaction rates following Eq.~(\ref{eq:reaction_ratesAB}. %For simplicity, we set $\chi_{A0}=\chi_{B0}=0$, 
%$\chi=\unit[6.5]{k_BT\mu m^{-3}}$, $\omega_A=\omega_B=\unit[0.5]{k_BT}$ and $\kappa_A=\kappa_B=\unit[2k_BT]{\mu m^{-2}}$. 
The parameters are given in Appendix~\ref{app:numerics}. (c): Single-molecule drift profiles for molecule $A$ and $B$ as given in Eq.~(\ref{eq:v_alpha}). The dashed line is the effective long-term drift $\tilde{v}$ given in Eq.~(\ref{eq:vtilde}). 
      The circles are mean local velocities for each enzyme conformation, as defined in Eq.~\ref{eq:mean}, 
      for $\Delta t=\unit[0.001]{s}$. 
      The gray dashed line is the effective drift following Eq.~\eqref{eq:veff} and the gray stars are local mean velocities (irrespective of the enzyme conformation), as defined in Eq.~\ref{eq:mean_irr}, for the longer time step $\Delta t=\unit[0.01]{s}$.}
			\label{fig:neq}
\end{figure}
During the time that typically elapses between two reaction events, the enzyme travels on average a distance of about $\ell_D=\sqrt{D/K}$ due to diffusion and $\ell_v=v/K$ due to the drift, where $K=K_++K_-$.
If the reaction rates vary weakly over these distances, ($K/\nabla K\gg \ell_D$)
the enzyme has time to relax to local equilibrium and it reaches a local
 probability $q_A(x,t)=K_-/(K_++K_-) $ and $q_B(x,t)=K_+/(K_++K_-)$ of being in state $A$ or $B$, respectively.   
These probabilities depend on composition because of the interactions captured by $\chi$ (see Ref.~\cite{Bauermann2022b}), %and can depend on position as well because of the space dependent profiles of substrate $S$ and product $Q$. 
%These probabilities
and, therefore also on position. 
%
% It is then possible to compute the effective drift of the enzyme and the enhanced diffusion. 
%due to the switching between different conformations.
%
%The probabilities are proportional to the average time spent in a conformation by the enzyme. 
On long time scales $\tau\gg K^{-1}$, the enzyme undergoes many reactions, switching many times between its two conformations. 
Observing the system on such time scales, we see effective dynamics, which
corresponds to averaging over the different conformational states of the enzyme \cite{pavliotis2008multiscale,Bo2017}.
Thus at long times, the enzyme moves following the effective Langevin equation
\begin{align}\label{eq:eff_lang}
    d{ x}={ v}^{\rm eff}+\nabla D^{\rm eff}+\sqrt{2D^{\rm eff}}d{w}_t
\end{align}
where ${ x}$ is the position of the enzyme (irrespective of its conformation), $w$ is Gaussian white noise with the same properties as in Eq.~(\ref{eq:langevin_x}) and the multiplicative noise follows Ito's discretization. For notational ease, we have restricted ourselves to the one-dimensional case, and discuss the multidimensional case in Appendix \ref{app:multi_multi}. 
% \begin{align}\label{eq:eff_lang}
%     d{\bf x}={\bf v}^{\rm eff}+\sqrt{2D^{\rm eff}}d{\VEC{w}}_t
% \end{align}
% where ${\bf x}$ is the position of the enzyme (irrespective of its conformation), ${\VEC{w}}$ is Gaussian white noise with the same properties as in Eq.~(\ref{eq:langevin_x}) and the multiplicative noise follows Ito's discretization.
Using a multiple-scale approach \cite{pavliotis2008multiscale,Bo2017}, following the method of Ref.~\cite{Aurell2017}, we obtain 
the long-term effective drift and diffusion coefficients for the case when the drift and diffusion coefficient vary on length scales larger than $\ell$.
The effective long-term drift reads
\begin{align}\label{eq:veff}
    v^{\mathrm{eff}} = \langle v\rangle + v_s\,,
\end{align}
% $v^{\mathrm{eff}} = \langle v\rangle + v_s$,
% \begin{equation}
% \label{eq:eff_drift}
% v^{\mathrm{eff}} = \langle v\rangle + v_s\,,
% \end{equation}
where
$
    \langle v\rangle=q_A v_A + q_B v_B
$
is the average drift, 
and $v_s$, containing corrections due to the spatial dependence of rates and drifts, is given in Eq.~(\ref{eq:eff_drift_s}).
The effective diffusion coefficient is $D^{\mathrm{eff}} = \langle D\rangle+D^{\mathrm{enh}}$
% \begin{align}
% \label{eq:eff_D}
% D^{\mathrm{eff}} &= \langle D\rangle + \frac{ q_Aq_B}{K_++K_-}\left(v_A-v_B\right)^2,
% % \\\nonumber
% %                  &=\langle D\rangle + \frac{ q_Aq_B}{K_++K_-}\left(m\nabla\lambda_\chi\right)^2\,,
% \end{align}
where the average diffusion coefficient is defined as
$
     \langle D\rangle\equiv q_AD_A+q_BD_B%\frac{ \left(K_+ D + K_- D^*\right)}{K_++K_-}
$.
When the diffusion coefficients do not depend on space, $\langle D\rangle=D$.
 $D^{\mathrm{enh}} =  q_Aq_B\left(v_A-v_B\right)^2/K$ 
is an enhancement to diffusion stemming from the additional noise caused by switching between different conformations.  When the diffusion coefficients of the two conformations of the enzyme are similar (as in the example we are considering, where $D_A=D_B$), because of the relation between the drifts and the gradients of the reaction rates \eqref{eq:vandK}, we expect the diffusion enhancement to be small where the conditions for the multiple-scale analysis to be accurate are satisfied: $K/\nabla K\gg \ell_D$, as shown in Eq.~\eqref{eq:Denh_ratio_new}.
\subsection{Numerical study of an enzyme in a nonequilibrium phase-separated mixture.}
When a system is driven out of equilibrium, it can display gradients in the volume fraction profiles within phases.
We provide an illustrative one-dimensional example where we maintain a phase-separating ternary mixture out of equilibrium by introducing sink and source terms at the system's boundaries.
The steady-state composition profile can be obtained numerically (see Appendix~\ref{app:numerics}). As shown in Fig.~\ref{fig:neq}(a),  the profile displays gradients imposed by the nonequilibrium boundaries. These gradients sustain steady volume fraction fluxes across the system, which are shown in Fig.~\ref{fig:neq}(b) together with the volume fraction reaction rates.
With the numerically obtained profiles, we can explicitly evaluate the drifts experienced by the two conformations of the enzyme given in Eq.~(\ref{eq:v_alpha}). 
We show the profiles of the single-molecule drifts in Fig.~\ref{fig:neq}(c), highlighting that an enzyme experiences a different drift depending on its conformation. On longer time scales, the gray dashed lines and symbols show that the enzyme experiences a net drift, which is in agreement with the prediction from the multiple-scale analysis \eqref{eq:veff}.

\section{Conclusion and discussion}
We have shown how labeling a fraction of molecules in a phase-separating system with chemical reactions can unveil single-molecule properties that are hidden when only bulk measurements are available. Considering the free energy of the system and taking an irreversible thermodynamics approach allowed us to derive the single-molecule drift, diffusion coefficient, and reaction rates, which govern the stochastic dynamics of interacting individual molecules undergoing chemical reactions. 
The interactions between the molecules determine the volume fraction profiles and can give rise to phase separation. These interactions also influence the chemical kinetics. Deriving the single-molecule dynamics has allowed us to study how such interactions and phase separation govern the fluctuations of single molecules.
We unveiled how kinetic parameters ({\it e.g.}, mobility) and thermodynamic ones ({\it e.g.}, inter-particle interactions) determine the motion of single molecules and their statistics. This opens the possibility of inferring these physical properties from the experimental recording of individual molecular trajectories. With recent advances in single-molecule localization microscopy \cite{Balzarotti2017NanometerFluxes}, sufficiently long tracks might be achievable, opening a window for the experimental measurement of the physical properties of small biomolecular condensates that have proven elusive for bulk measurements. \\ %providing an exciting test bed for our theory in biological systems.\\  
Focusing on nonequilibrium conditions we showed the emergence of a systematic directed motion at the level of single molecules due to the combination of concentration gradients with interactions. This observation provides an additional possible mechanism for the chemotaxis of active enzymes which was reported in different conditions \cite{Agudo-Canalejo2018PhoresisChemotaxis,Agudo-Canalejo2018EnhancedNanoscale,Agudo-Canalejo2020CooperativelyProteins,Feng2020EnhancedEnzymes}.
Previous studies showed that when a single molecule randomly switches between states that experience different drifts, there is an additional dispersion in the position which effectively enhances diffusion. This mechanism is akin to that of Taylor dispersion for particles diffusing in a shear flow~\cite{Taylor1953}, which was also discussed for particles switching between different diffusive states ~\cite{Mysels1956,Aurell2016a,Aurell2017,Kahlen2017,Pietzonka2016,Agudo-Canalejo2020DiffusionEnzymes} and for particles near charged surfaces~\cite{Vilquin2023NanoparticleBoundary}.
However, we find that this enhancement is small when the drifts are not due to external forces but only due to the intermolecular interactions, which also govern the rates and the free energy of Eq.~\eqref{eq:free_energy_compactAB}.\\
Our findings provide additional tools to investigate how phase separation impacts the fluctuations of molecules undergoing chemical reactions \cite{Klosin2020,Deviri2021,Zechner2025ConcentrationSystems}, which is an interesting direction for future research. It will also be interesting to consider the impact of interface resistance \cite{Zhang2024TheCondensatesb,Hubatsch2024TransportPhases} for the motion of individual molecules in chemically active condensates.
\subsection*{Acknowledgements}
The authors wish to thank Omar Adame Arana, Jonathan Bauermann, and Christoph A. Weber for helpful discussions, and Jaime Agudo-Canalejo for suggesting references.

\onecolumngrid
\appendix

\section{Free energy, exchange chemical potentials and reaction rates for a ternary mixture}%: explicit expressions}
The free energy density for a ternary mixture reads
% \begin{align}\label{eq:free_energy_compactAB}
% f =&\frac{k_BT}{\nu}(\phi_A\ln\phi_A  + 
% \phi_B\ln\phi_B+
% \frac{\nu}{\nu_0}\phi_0\ln\phi_0)\\\nonumber
% &+\chi_{AB}\phi_A\phi_B +
% \chi_{A0}\phi_A\phi_0 +
% \chi_{B0}\phi_B\phi_0 \\\nonumber
% &+\omega^0_A\phi_A\,+
% \omega^0_B\phi_B\,+
% \omega^0_0\phi_0\, ,
% \end{align}
\begin{align}\label{eq:free_energy_compactAB}
f =\frac{k_BT}{\nu}(\phi_A\ln\phi_A  + 
\phi_B\ln\phi_B+
\frac{\nu}{\nu_0}\phi_0\ln\phi_0)+\chi_{AB}\phi_A\phi_B +
\chi_{A0}\phi_A\phi_0 +
\chi_{B0}\phi_B\phi_0 +\omega^0_A\phi_A\,+
\omega^0_B\phi_B\,+
\omega^0_0\phi_0\, ,
\end{align}
where $\phi_0=1-\phi_A-\phi_B$ is implied, $\chi_{\alpha\beta}$ denotes the interaction parameter, $\nu_0$ is the molecular volume of the solvent and $\omega^0_\alpha$ are the internal free energy densities for the components. 
The exchange chemical potential $\mu_\alpha=\nu_\alpha \delta F/\delta \phi_\alpha$ is computed from the free energy of the ternary mixture which is defined in  Eqs.~\eqref{eq:free_energyAB}, \eqref{eq:free_energy_compactAB}, and reads
\begin{align}\label{eq:chem_potAB}
\mu_A&=k_BT\ln  \left( \phi_0^{-\frac{\nu}{\nu_0}} \phi_A \right)+\nu \left[
-2\chi_{A0}\phi_A +
(\chi_{AB}-\chi_{A0}-\chi_{B0}) \phi_B 
\right] + \omega_A - \kappa_A \nabla^2 \phi_A \\\nonumber
\mu_B&=k_BT \ln  \left( \phi_0^{-\frac{\nu}{\nu_0}} \phi_B \right)+\nu \left[
-2\chi_{B0}\phi_B +
(\chi_{AB}-\chi_{A0}-\chi_{B0}) \phi_A 
\right] + \omega_B - \kappa_B \nabla^2 \phi_B
\, ,
\end{align}
where  $\omega_\alpha$ are composition-independent reference chemical potentials:
% given in Eq.~\eqref{eq:reference_muAB}.

% The composition-independent reference chemical potentials follow from taking the derivative of the free energy given in Eq.~\eqref{eq:free_energyAB} and read
\begin{align}\label{eq:reference_muAB}
\omega_A =& k_BT\left(1- \frac{\nu}{\nu_0}\right) + \nu\chi_{A0} + \nu\left( \omega_A^0-  \omega_0^0 \right)\\\nonumber
\omega_B =& k_BT\left(1- \frac{\nu}{\nu_0}\right) + \nu\chi_{B0} + \nu\left( \omega_B^0-  \omega_0^0 \right)\,,
\end{align}
which we assume to be constant in space.
We consider systems obeying detailed balance so that the ratio of the reaction rates is set by the differences in chemical potentials between reagent and product species, as defined in Eq.~\eqref{eq:detailed_balance} in the main text. The reaction ratio then reads
\begin{align}\label{eq:r_ratio}
    \frac{r_+}{r_-}=e^{\frac{\mu_A-\mu_B}{k_B T}}=\frac{\phi_A}{\phi_B}\exp{\left[\frac{1}{k_BT}\left(\nu\lambda_\chi(\phi_A,\phi_B)+\omega_A-\omega_B
    +\lambda_\kappa
    \right)
    \right]}
\end{align}
where we have defined
\begin{align}\label{eq:lambda_chi}
    \lambda_\chi (\phi_A,\phi_B)=\chi_{AB}(\phi_B-\phi_A)+\left(\chi_{B0}-\chi_{A0}\right)(\phi_A+\phi_B)\,,
\end{align}
as the difference in chemical potentials due to the interactions $\chi$ and
\begin{align}\label{eq:lambda_kappa}
    \lambda_\kappa (\phi_A,\phi_B)=
    - \kappa_A \nabla^2 \phi_A+ \kappa_B \nabla^2 \phi_B\,.
\end{align}
The individual rates can be obtained from the definitions of Eq.~\eqref{eq:reaction_ratesAB}, and that of the chemical potentials \eqref{eq:chem_potAB}
\begin{align}\label{eq:r_explicit}
   r_+&=k\frac{\phi_A}{\phi_0}\exp{\left[\frac{1}{k_BT}\left(\nu \left(
-2\chi_{A0}\phi_A +
\tilde{\chi} \phi_B 
\right)+\omega_A
    - \kappa_A \nabla^2 \phi_A
    \right)
    \right]}\\\nonumber
    r_-&=k\frac{\phi_B}{\phi_0}\exp{\left[\frac{1}{k_BT}\left(\nu \left(
-2\chi_{B0}\phi_B +\tilde{\chi}\phi_A 
\right)+\omega_B
    - \kappa_B \nabla^2 \phi_B
    \right)
    \right]}\,.
\end{align}
where we have defined
\begin{align}\label{eq:chitilde}
    \tilde{\chi}\equiv\chi_{AB}-\chi_{B0}-\chi_{A0}\,.
\end{align}

The fluxes of volume fractions are driven by gradients in the chemical potential [See Eq.~(\ref{eq:j})]. Using the explicit expression for the mobility provided in Eq.~(\ref{eq:MAB}), the fluxes read
\begin{align}\label{eq:jAB_app}
    \boldsymbol{j}_A=&-\left[m_{A0}\phi_A\phi_0\nabla\mu_A+m_{AB}\phi_A\phi_B\nabla(\mu_A-\mu_B)
    \right]\\\nonumber
    \boldsymbol{j}_B=&-\left[m_{B0}\phi_B\phi_0\nabla\mu_B+m_{AB}\phi_A\phi_B\nabla(\mu_B-\mu_A)\right]\,.
\end{align}
These expressions for the fluxes can be used to define the collective diffusion matrix $D^{\text{col}}$ via
\begin{align}\label{eq:D_col}
     \boldsymbol{j}_\alpha=-\sum_\beta D^{\text{col}}_{\alpha\beta}\nabla \phi_\beta\,.
\end{align}
\section{Labeled molecules}\label{app:labeled}
Labeling molecules changes the entropy of the mixtures since our ability to distinguish them increases the number of possible configurations. We assume that labeling does not modify the interactions between molecules or their internal energy. With these assumptions, the free energy density reads:
\begin{align}\label{eq:free_energy_compactAB_labeled}
\tilde{f} =&\frac{k_BT}{\nu}(\phi_A^*\ln\phi_A^*  +
\overline{\phi}_A\ln\overline{\phi}_A+
\phi_B^*\ln\phi_B^*+
\overline{\phi}_B\ln\overline{\phi}_B+
\frac{\nu}{\nu_0}\phi_0\ln\phi_0)\\\nonumber
&+\chi_{AB}\phi_A\phi_B +
\chi_{A0}\phi_A\phi_0 +
\chi_{B0}\phi_B\phi_0 +\omega^0_A\phi_A\,+
\omega^0_B\phi_B\,+
\omega^0_0\phi_0\, .
\end{align}

%\subsection{Reaction dynamics of labeled and unlabled molecules}

\subsection{Diffusion flux of labeled and unlabeled molecules}
% Our assumptions that labeling does not affect the physical properties of the molecules implies that the only modification to the free energy due to the labeling is in the  the mixing entropy contribution, since we now have two additional species. 
Following linear irreversible thermodynamics, fluxes are driven by gradients in chemical potentials as in Eq.~{\ref{eq:j}},
with the sum that now runs over 4 components $A^*,\,\overline{A},\,B^*,\,\overline{B}$.
Concerning kinetic coefficients, the mobilities of labeled and unlabeled molecules are unchanged. 
However, due to the labeling, we now have to specify how labeled and unlabeled molecules of the same kind ({\it e.g.}, $A^*$ and $\overline{A}$ or $B^*$ and $\overline{B}$) exchange position. This information is encoded in the coefficients $m_{AA}$ and $m_{BB}$, which were not necessary to describe the evolution of the total volume fraction $\phi_A$ and  $\phi_B$.
This is evident when we consider the mobility matrix for a multicomponent mixture
$M_{\alpha\beta}$ following Refs.~\cite{Kramer1984,Mao2020,Bo2021a,Cotton2022}:
\begin{gather}\label{eq:MAB_lab}
M_{ii} =   m_{i0} \phi_\alpha\phi_0 +
\sum_{j\neq i}^n  m_{\alpha\beta} \phi_\alpha \phi_\beta \\\nonumber
M_{\alpha\beta} = -m_{\alpha\beta} \phi_\alpha \phi_\beta, \;\;\; \forall i \neq j \,.
\end{gather}
For the ternary mixtures with labels that we are considering, we have
\begin{align}\label{eq:MAB_labeled}
     M=
     \left( \begin{array}{c c c c c c c}
(m_{A0}\phi_A^*\phi_0+ 
&\qquad&
&\qquad&
&\qquad&
\\
m_{AA}\phi_A^*\overline{\phi}_A+
&\qquad&-m_{AA}\phi_A^*\,\overline{\phi}_A
&\qquad&-m_{AB}\phi_A^* \,\phi_B^*	
&\qquad& -m_{AB}\phi_A^* \,\overline{\phi}_B	
\\
m_{AB} \phi_A^*\phi_B)
&\qquad&
&\qquad&
&\qquad&
\\[6pt]
&\qquad& (m_{A0}\overline{\phi}_A\phi_0+
&\qquad&
&\qquad&
\\
-m_{AA}\phi_A^*\,\overline{\phi}_A
&\qquad& m_{AA}\overline{\phi}_A\phi^*_A+
&\qquad&-m_{AB}\overline{\phi}_A\,\phi_B^*
&\qquad& -m_{AB}\overline{\phi}_A\,\overline{\phi}_B\\
&\qquad&m_{AB} \overline{\phi}_A\phi_B)
&\qquad&
&\qquad&
\\[6pt]
&\qquad&
&\qquad&(m_{B0}\phi_B^*\phi_0+
&\qquad&\\
   -m_{AB}\phi_A^*\, \phi_B^*	
 &\qquad& -m_{AB}\overline{\phi}_A\,\phi_B^*
 &\qquad& m_{BB}\phi_B^*\overline{\phi}_B+ 
 &\qquad& -m_{BB}\phi_B^*\,\overline{\phi}_B\\
 &\qquad&
 &\qquad&m_{AB} \phi_B^*\phi_A)
 &\qquad&\\
\\[6pt]
&\qquad&
&\qquad&
&\qquad&(m_{B0}\overline{\phi}_B\phi_0+\\
   -m_{AB}\phi_A^* \overline{\phi}_B	
 &\qquad& -m_{AB}\overline{\phi}_A\,\overline{\phi}_B
 &\qquad& -m_{BB}\phi_B^*\,\overline{\phi}_B
 &\qquad& m_{BB}\overline{\phi}_B\phi_B^* + \\
 &\qquad&
 &\qquad&
 &\qquad&m_{AB} \overline{\phi}_B\phi_A)
\end{array}
\right)
 \end{align}
 Plugging this expression for the mobility~\eqref{eq:MAB_labeled} and the chemical potentials obtained from the free energy of the labeled multicomponent mixture~\eqref{eq:free_energy_compactAB_labeled} [given in Eq.~\eqref{eq:chemical_potential_AB_lab}] into the equation for the flux Eq.~{\ref{eq:j}}, we obtain the fluxes of labeled molecules of species $A$ and $B$ 
\begin{align}\label{eq:j_multi}
    \boldsymbol{j}^*_{A} = \frac{\phi_A^*}{\phi_A}\boldsymbol{j}_A+D_A\left(-\nabla \phi_A^*+ \phi_A^*\frac{\nabla\phi_A}{\phi_A}\right)\,,\\
    \boldsymbol{j}^*_{B} = \frac{\phi_B^*}{\phi_B}\boldsymbol{j}_B+D_B\left(-\nabla \phi_B^*+ \phi_B^*\frac{\nabla\phi_B}{\phi_B}\right)\,,
\end{align}
which coincide with the expression for the diffusion flux of labeled molecules derived for the case without chemical reactions, in \cite{Bo2021a}.
As reported in Eq.~\eqref{eq:DAB} in the main text, the single-molecule diffusion coefficients are
 \begin{align}\label{eq:DA_app}
D_A(x,t)=&k_BT\left(m_{A 0}\phi_0+m_{A A}\phi_A+m_{A B}\phi_B\right)\,,
\\
\nonumber
D_B(x,t)=&k_BT\left(m_{B 0}\phi_0+m_{A B}\phi_A+m_{B B}\phi_B\right)
\,.
%g_B=&1-\xi_{B B}\phi_B-\xi_{BA}\phi_A
\end{align}
 \subsubsection{Labeling conserved quantities}

In case the labeling does not distinguish between species $A$ and $B$, the dynamics of 
 the  labeled molecules $\phi^*=\phi_A^*+\phi_B^*$ follows by summing the two equations in \eqref{eq:effective_rate_AB}:
\begin{align}
\partial_\mathrm{t}\phi^*
=&%k_\mathrm{B}T \,
\nabla\cdot\left[D_{A}\left(\nabla \phi_A^*-\phi_A^*\frac{\nabla\phi_A}{\phi_A}\right)-\phi_A^*\frac{\boldsymbol{j}_A}{\phi_A}\right]
+
\nabla\cdot\left[D_{B}\left(\nabla \phi_B^*-\phi_B^*\frac{\nabla\phi_B}{\phi_B}\right)-\phi_B^*\frac{\boldsymbol{j}_B}{\phi_B}\right]\, .
\label{eq:effective_rate_conserved}
\end{align}
Importantly, this dynamics depends also on $\phi_A^*$ and $\phi_B^*$, which cannot be distinguished and observed individually.
 \subsubsection{Alternative expression for the multicomponent dynamics of labeled molecules}

To highlight the role of nonequilibrium, and introducing $r=r_+-r_-$  we can rewrite Eq.~(\ref{eq:effective_rate_AB}) as
\begin{align}\label{eq:effective_rate_AB_neq}
\partial_\mathrm{t}\phi_A^*
=&%k_\mathrm{B}T \,
\nabla\cdot\left[D_{A}\left(\nabla \phi_A^*-\phi_A^*\frac{\nabla\phi_A}{\phi_A}\right)-\phi_A^*\frac{\boldsymbol{j}_A}{\phi_A}\right]
-\frac{r_-}{\phi_B}\left(\frac{\phi_B}{\phi_A}\phi_A^*-\phi_B^*\right)- r\frac{\phi_A^*}{\phi_A}\, ,\\\nonumber
% \phi_A^*\frac{\boldsymbol{j}_A}{\phi_A}\right]-\frac{r^_{A\to B}}{\phi_A}\phi_A^*+\frac{r^-_{A\to B}}{\phi_B}\phi_B^*\, ,
\partial_\mathrm{t}\phi_B^*
=&%k_\mathrm{B}T \,
\nabla\cdot\left[D_{B}\left(\nabla \phi_B^*-\phi_B^*\frac{\nabla\phi_B}{\phi_B}\right)-\phi_B^*\frac{\boldsymbol{j}_B}{\phi_B}\right]
+\frac{r_-}{\phi_B}\left(\frac{\phi_B}{\phi_A}\phi_A^*-\phi_B^*\right)+r\frac{\phi_A^*}{\phi_A}\, .
% \phi_B^*\frac{\boldsymbol{j}_B}{\phi_B}\right]-\frac{r^-_{A\to B}}{\phi_B}\phi_B^*+\frac{r^+_{A\to B}}{\phi_A}\phi_A^*\, .
%\label{eq:effective_rate_B}
\end{align}

\subsubsection{Labeled molecules in an equilibrium mixture}
When the total mixture is at equilibrium, the fluxes vanish $\boldsymbol{j}_\alpha=0$, $\tilde{r}=r_+=r_-$ and the expression simplifies to
\begin{align}\label{eq:effective_rate_AB_eq}
\partial_\mathrm{t}\phi_A^*
=&%k_\mathrm{B}T \,
\nabla\cdot\left[D_{A}\left(\nabla \phi_A^*-\phi_A^*\frac{\nabla W_A^{eq}}{k_BT}\right)\right]
-\frac{\tilde{r}}{\phi_B^{eq}}\left(\mathcal{K}\phi_A^*-\phi_B^*\right)\, ,\\\nonumber
% \phi_A^*\frac{\boldsymbol{j}_A}{\phi_A}\right]-\frac{r^_{A\to B}}{\phi_A}\phi_A^*+\frac{r^-_{A\to B}}{\phi_B}\phi_B^*\, ,
\partial_\mathrm{t}\phi_B^*
=&%k_\mathrm{B}T \,
\nabla\cdot\left[D_{B}\left(\nabla \phi_B^*-\phi_B^*\frac{\nabla W^{eq}_B}{k_BT}\right)\right]
+\frac{\tilde{r}}{\phi_B^{eq}}\left(\mathcal{K}\phi_A^*-\phi_B^*\right)\, ,
% \phi_B^*\frac{\boldsymbol{j}_B}{\phi_B}\right]-\frac{r^-_{A\to B}}{\phi_B}\phi_B^*+\frac{r^+_{A\to B}}{\phi_A}\phi_A^*\, ,
%\label{eq:effective_rate_B}
\end{align}

where $\phi_\alpha^{eq}$ is the equilibrium volume fraction of species $\alpha$ and $W_\alpha^{eq}=-k_BT\ln\phi_\alpha^{eq}$, which in a phase-separating system depends on position. Furthermore, $\mathcal{K}=\phi_B^{eq}/\phi_A^{eq}$
% \begin{align}\label{eq:Keq}
%     K=\frac{\phi_B^{eq}}{\phi_A^{eq}}%=\phi_B^{eq}/\phi_A^{eq}
% \end{align}
 describes the chemical equilibrium and is in general different in coexisting phases \cite{Bauermann2022b}.
 \subsection{Single-molecule reaction rates}
 The single-molecule rates can be obtaining plugging into their expression~\eqref{eq:K}, the one for the collective reaction rates~\eqref{eq:r_explicit}. 
 For equal molecular volumes, $\nu=\nu_0$, these read
%  \begin{align}\label{eq:K_explicit}
%    K_+&=\frac{k}{\phi_0}\exp{\left[\frac{1}{k_BT}\left(\nu \left(
% -2\chi_{A0}\phi_A +
% (\chi_{AB}-\chi_{A0}-\chi_{B0}) \phi_B 
% \right)+\omega_A
%     - \kappa_A \nabla^2 \phi_A
%     \right)
%     \right]}\\\nonumber
%     K_-&=\frac{k}{\phi_0}\exp{\left[\frac{1}{k_BT}\left(\nu \left(
% -2\chi_{B0}\phi_B +
% (\chi_{AB}-\chi_{B0}-\chi_{A0}) \phi_A 
% \right)+\omega_B
%     - \kappa_B \nabla^2 \phi_B
%     \right)
%     \right]}\,.
% \end{align}
\begin{align}\label{eq:K_explicit}
   K_+&=\frac{k}{\phi_0}\exp{\left[\frac{1}{k_BT}\left(\nu \left(
-2\chi_{A0}\phi_A +
\tilde{\chi} \phi_B 
\right)+\omega_A
    - \kappa_A \nabla^2 \phi_A
    \right)
    \right]}\\\nonumber
    K_-&=\frac{k}{\phi_0}\exp{\left[\frac{1}{k_BT}\left(\nu \left(
-2\chi_{B0}\phi_B +\tilde{\chi}\phi_A 
\right)+\omega_B
    - \kappa_B \nabla^2 \phi_B
    \right)
    \right]}\,.
\end{align}
% where we have defined
% \begin{align}\label{eq:chitilde}
%     \tilde{\chi}\equiv\chi_{AB}-\chi_{B0}-\chi_{A0}
% \end{align}
\subsection{Single-molecule drift}
To compute the single-molecule drift we need to plug 
the expression for the flux, Eq.~\eqref{eq:jAB_app}, and the one for the chemical potential, Eq.~\eqref{eq:chem_potAB}, into the definition of the drift~\eqref{eq:vAB}.
For a one-dimensional case with a constant single-molecule diffusion coefficient $D=mk_BT$ (independent of composition and position) with molecules having the same molecular volume of the solvent $\nu=\nu_0$,
the  expression for the single-molecule drift reads
% \begin{align}\label{eq:drifts}
% v_A=&-m\nu \nabla\phi_A\big[  -2\chi_{A0}\phi_0 + \phi_B (-\chi_{AB}-\chi_{A0}+\chi_{B0})\big]+\\\nonumber
% &-m\nu\nabla\phi_B \big[\phi_0\left(\chi_{AB}-\chi_{A0}-\chi_{B0}\right) +
% \phi_B\left(\chi_{AB}-\chi_{A0}+\chi_{B0}\right)\big]+\mathcal{C}_A\\\nonumber
%     v_B=&-m\nu \nabla \phi_A \big[\phi_0\left(\chi_{AB}-\chi_{A0}-\chi_{B0}\right)+ \phi_A\left(\chi_{AB}+\chi_{A0}-\chi_{B0}\right)\big]+\\\nonumber
% &-m\nu\nabla\phi_B\big[  -2\chi_{B0}\phi_0 + \phi_A (-\chi_{AB}+\chi_{A0}-\chi_{B0})\big]+\mathcal{C}_B\,,
% \end{align}
% \begin{align}\label{eq:drifts}
% v_A=&-m\nu \nabla\phi_A\big[  -2\chi_{A0}\phi_0 - \phi_B( \tilde{\chi}+2\chi_{A0})\big]+\\\nonumber
% &-m\nu\nabla\phi_B \big[\tilde{\chi}\phi_0+
% \phi_B( \tilde{\chi}+2\chi_{B0})\big]+\mathcal{C}_A\\\nonumber
%     v_B=&-m\nu \nabla \phi_A \big[\tilde{\chi}\phi_0+ \phi_A\left( \tilde{\chi}+2\chi_{A0}\right)\big]+\\\nonumber
% &-m\nu\nabla\phi_B\big[  -2\chi_{B0}\phi_0 -\phi_A (\tilde{\chi}+2\chi_{B0})\big]+\mathcal{C}_B\,,
% \end{align}
%
%
%
\begin{align}\label{eq:drifts}
v_A=&m\nu \left\{\nabla\phi_A\big[\tilde{\chi}\phi_B+  2\chi_{A0}(1-\phi_A)   \big]
-\nabla\phi_B \big[\tilde{\chi}(1-\phi_A)+
2\chi_{B0}\phi_B\big]+\left[\kappa_A\left(1-\phi_A\right)\nabla \nabla^2\phi_A-
    \kappa_B\phi_B\nabla \nabla^2\phi_B\right]\right\}
%+m\left[\phi_B\nabla\omega_B-\left(1-\phi_A\right)\nabla\omega_A\right]
\\
\nonumber
    v_B=&m\nu \left\{\nabla\phi_B\big[\tilde{\chi}\phi_A +  2\chi_{B0}(1-\phi_B) \big]
    -\nabla \phi_A \big[\tilde{\chi}(1-\phi_B)+ 2\chi_{A0}\phi_A\big]+\left[\kappa_B\left(1-\phi_B\right)\nabla \nabla^2\phi_B-
    \kappa_A\phi_A\nabla \nabla^2\phi_A\right]\right\}
   % +m\left[\phi_A\nabla\omega_A-\left(1-\phi_B\right)\nabla\omega_B\right]
   \,.
\end{align}
%
% where
% \begin{align}\label{eq:curva}
%     \mathcal{C}_A&=m\nu\left[\kappa_A\left(1-\phi_A\right)\nabla \nabla^2\phi_A-
%     \kappa_B\phi_B\nabla \nabla^2\phi_B\right]\\\nonumber
%     \mathcal{C}_B&=m\nu\left[\kappa_B\left(1-\phi_B\right)\nabla \nabla^2\phi_B-
%     \kappa_A\phi_A\nabla \nabla^2\phi_A\right]\,,
% \end{align}
and $\tilde{\chi}$ is defined in Eq.~\eqref{eq:chitilde}.
One can rewrite the drifts as in Eq.~\eqref{eq:v_alpha} in the main text
\begin{align}\label{eq:v_alpha_app}
    v_\alpha=m\left[\nu\tilde{\chi}\sum_\beta g_{\alpha\beta}\nabla\phi_\beta+\tilde{\kappa}\sum_\beta c_{\alpha\beta}\nabla\nabla^2\phi_\beta\right]
\end{align}
by introducing $\tilde{\kappa}=(\kappa_A+\kappa_B)/2$,
% \begin{align}\label{eq:drifts_rw}
% v_\alpha=m\nu\tilde{\chi}g_{\alpha\beta}\nabla \phi_\beta%+m\tilde{g}_{\alpha\beta}\nabla \omega_\beta
% +\mathcal{C}_\alpha
% \,,
% \end{align}
% where 
\begin{align}\label{eq:g}
g_{\alpha\beta}=
\begin{cases}
-\phi_0+(1-\phi_\alpha)\frac{\tilde{\chi}+2\chi_{\alpha0}}{\tilde{\chi}}& \mbox{if} \qquad \alpha=\beta\\
-\phi_0-\phi_\beta\frac{\tilde{\chi}+2\chi_{\beta0}}{\tilde{\chi}}
     %-(1-\phi_\alpha)-2\frac{\chi_{\beta0}}{\tilde{\chi}}\phi_\beta
     & \mbox{if}\qquad  \alpha\neq\beta
     \end{cases}\,,
\end{align}
and 
\begin{align}\label{eq:c}
c_{\alpha\beta}=
\begin{cases}
(1-\phi_\alpha)\frac{\kappa_{\alpha}}{\tilde{\kappa}}& \mbox{if} \qquad \alpha=\beta\\
-\phi_\beta\frac{\kappa_{\beta}}{\tilde{\kappa}}
     %-(1-\phi_\alpha)-2\frac{\chi_{\beta0}}{\tilde{\chi}}\phi_\beta
     & \mbox{if}\qquad  \alpha\neq\beta
     \end{cases}\,.
\end{align}
When all the mobilities are equal ($m_{\alpha\beta}=m$),  $D_A=D_B=k_BTm$, using the 
the expression for the flux, Eq.~\eqref{eq:jAB_app}, and the definition of the drift~\eqref{eq:vAB} we can compute the difference between the single-molecule drifts 
\begin{align}\label{eq:vA-vB_mu_app}
    v_A-v_B=m\nabla\left(k_BT\ln{\frac{\phi_A}{\phi_B}}-\mu_A+\mu_B\right)\,.
\end{align}
For generic mobilities, one gets
\begin{align}\label{eq:vA-vB_mu_app_gen}
    v_A-v_B=&-D_A\nabla\left(\mu_A-k_BT\ln{\phi_A}\right)+D_B\left(\mu_B -k_BT\ln{\phi_B}\right)\\\nonumber
    &+\phi_A\left(m_{AA}-m_{AB}\right)\nabla \mu_A-\phi_B\left(m_{BB}-m_{AB}\right)\nabla \mu_B
\,.
\end{align}
\section{Multiple-scale techniques to compute the long-term effective drift and enhanced diffusivity}\label{app:multiscale}
To derive the long-term effective drift we use a multiple-scale technique \cite{pavliotis2008multiscale,Bo2017} and follow Ref~\cite{Aurell2017}. For simplicity, we consider a one-dimensional case.
We start by rewriting Eq.~\eqref{eq:FPAB} as 
\begin{eqnarray}\label{eq:fp_mat}
\frac{\partial}{\partial t}
\left(
 \begin{array}{c }
P_A(x)	 \\
 P_B(x)
\end{array}\right)
&=&\left( \begin{array}{c c }
-K_++{\cal L}^\dagger_+& K_-	 \\
K_+& -K_-	+{\cal L}^\dagger_-
\end{array}
\right)
\left( \begin{array}{c }
P_A(x)	 \\
 P_B(x)
\end{array}
\right)=
(\TENSOR{K}+{\cal L}^\dagger)\left( \begin{array}{c }
P_A(x)	 \\
 P_B(x)
\end{array}\right)
\end{eqnarray}
where
\begin{subequations}
\begin{eqnarray}\label{eq:L}
{\cal L}^\dagger_\alpha&=&\frac{\partial}{\partial x}\left( -v_\alpha+D_\alpha\frac{\partial}{\partial x} \right) 
%\\
%\label{eq:master}
%\TENSOR{K}\left( \begin{array}{c }
%P_{+}	 \\
%P_{-}
%\end{array}\right) &=& 
%\left( \begin{array}{c c }
%-K_+& K_-	 \\
%K_+& -K_-	
%\end{array}
%\right)
%\left( \begin{array}{c }
%P_{+}	 \\
% P_{-}
%\end{array}
%\right)
\end{eqnarray}
and %$\TENSOR{K}$ is defined as %in Eq.~\eqref{eq:master}.
\begin{align}
\TENSOR{K}=    \left( \begin{array}{c c }
-K_+& K_-	 \\
K_+& -K_-	
\end{array}
\right)\,.
\end{align}
\end{subequations}
The main idea behind the multiple-scale approach is to introduce two additional time scales $\vartheta$ and $\tau$ identifying intermediate and long time scales and one large length scale $X$. We will derive the effective dynamics on these scales $\tau\gg K^{-1}$ and $X\gg\ell$ (note that in the main text we denote $X$ as $\mathcal{L}_D$.
The time scales are defined by the scaling 
\begin{subequations}\label{eq:scales}
\begin{equation}
\label{eq:tscales}
\theta = \epsilon^0 t
\, , \quad
\vartheta = \epsilon^{1} t
\, , \quad
\tau = \epsilon^{2} t
\, ,
\end{equation} 
where we have also introduced the notation $\theta$ which refers to the original time scale of the problem. For the spatial scales we have
\begin{equation}
\label{eq:xalphascales}
\tilde{ x } = \epsilon^0  x 
\, , \quad
X  = \epsilon^{1}  x 
\,. 
\end{equation}
\end{subequations}
 This choice is motivated by the fact that we are looking for effective diffusive dynamics on large scales. We treat these variables as independent and allow the probability to be a function of all of them $\boldsymbol{P}(\theta,\vartheta,\tau,\tilde{ x }, X )$. We assume that the profile $\phi$ varies appreciably only on the large length scale $X$ so that the rates $K$ and the drift coefficients $v_\alpha$ only depend on $X$. This assumption may be difficult to satisfy near phase boundaries where the volume fraction profile may sharply change.\\ 
The first time scale (captured by the variable $\theta$) is associated with the chemical reactions and is of order $K^{-1}$. On this time scale, the particle moves on average a distance of about $\ell_{v_\alpha}=K^{-1} v_\alpha$ with a scatter of $\ell_{D_\alpha}=\sqrt{D_\alpha K^{-1}}$.
These lengths identify the first spatial scale $\tilde{ x } = \epsilon^0  x$. 
%; the instantaneous diffusion coefficient is hence of order unity in the variables $\theta$ and $\tilde{x}$.
%We are looking for the effective diffusive motion on the large spatial scale $X$ on the slow time scale $\tau$. 
% The second time $\vartheta$ in (\ref{eq:tscales}) is the time it takes for the particle to
% traverse the large spatial scale moving steadily with the mean velocity $v_\alpha$.
% ; in the present analysis this will be only needed
% in an intermediate technical step, and does not by itself generate new physical effects.
% With these definitions, $X $ is the the large scale space variable which is of order one only for very large $ x $.
% We are interested in finding the effective dynamics in $X$.
The small scale variable $\tilde{ x }$ %and $\tilde{ a }$
 corresponds to the original variable $ x $ %and $ a $
but is restricted to small scales by imposing periodic
boundary conditions for $P$ over the typical length covered during the relaxation time. 
Treating all the new variables as independent, the time and spatial derivatives in Eq.~\eqref{eq:FPAB} and \eqref{eq:fp_mat} can now be expressed as

\begin{eqnarray}
\label{eq:derivatives}
\frac{\partial}{\partial t}
&=& \frac{\partial}{\partial\theta} + \epsilon \frac{\partial}{\partial\vartheta} + \epsilon^2 \frac{\partial}{\partial\tau}
\, , \\\nonumber
\frac{\partial}{\partial x_\alpha}
&=& \frac{\partial}{\partial\tilde{x}_\alpha} + \epsilon \frac{\partial}{\partial X_\alpha}
\, ,
\end{eqnarray}
where we have used the scaling \eqref{eq:scales} and applied the chain rule.
The transition matrix
$\TENSOR{K}$ 
remains unchanged.
We  treat $\epsilon$ as a small perturbative
parameter and expand $P$ in powers of $\epsilon$,
\begin{equation}
\label{eq:pexpansion}
\boldsymbol{P}= \boldsymbol{P}^{(0)} + \epsilon \boldsymbol{P}^{(1)} + \epsilon^2 \boldsymbol{P}^{(2)} + \ldots
\, ,
\end{equation}
where all $\boldsymbol{P}(\theta,\vartheta,\tau,\tilde{ x }, X )$ depend explicitly 
on the various variables.
In these variables, $\boldsymbol{P}^{(0)}$ is normalized to one, while all other
$\boldsymbol{P}^{(i)}$ with $i>0$ are normalized to zero.
Plugging \eqref{eq:derivatives} and \eqref{eq:pexpansion} into
\eqref{eq:fp_mat}, and collecting terms of equal powers in
$\epsilon$ we obtain a
hierarchy of equations. At order $\epsilon^0$, $\epsilon^1$ and $\epsilon^2$ we have
\begin{subequations}
\label{eq:hierarchy}
\begin{eqnarray}
\frac{\partial \boldsymbol{P}^{(0)}}{\partial\theta} - \left( \tilde{\OP{L}}^{\dagger} +  \TENSOR{K}  \right) \boldsymbol{P}^{(0)} & = & 0
%\, ,
\label{eq:hierarchy0}
\\
\frac{\partial \boldsymbol{P}^{(1)}}{\partial\theta} - \left( \tilde{\OP{L}}^{\dagger} +  \TENSOR{K}  \right) \boldsymbol{P}^{(1)} & = &
- \frac{\partial \boldsymbol{P}^{(0)}}{\partial\vartheta}
- \frac{\partial}{\partial X} \left( \begin{array}{c }
v_AP^{(0)}_A	 \\
v_BP^{(0)}_B
\end{array}
\right)
+  \frac{\partial}{\partial \tilde{x}} \left( \begin{array}{c }
D_A\frac{\partial}{\partial X}P^{(0)}_A	 \\
D_B\frac{\partial}{\partial X}P^{(0)}_B
\end{array}
\right)
+  \frac{\partial}{\partial X} \left( \begin{array}{c }
D_A\frac{\partial}{\partial \tilde{x}}P^{(0)}_A	 \\
D_B\frac{\partial}{\partial \tilde{x}}P^{(0)}_B
\end{array}
\right)
%\, ,
\label{eq:hierarchy1}
\\
\frac{\partial \boldsymbol{P}^{(2)}}{\partial\theta} - \left( \tilde{\OP{L}}^{\dagger} +  \TENSOR{K}  \right) \boldsymbol{P}^{(2)} & = &
- \frac{\partial \boldsymbol{P}^{(0)}}{\partial\tau} -\frac{\partial \boldsymbol{P}^{(1)}}{\partial\vartheta}
- \frac{\partial}{\partial X} \left( \begin{array}{c }
v_AP^{(1)}_A	 \\
v_B P^{(1)}_B
\end{array}
\right)\label{eq:hierarchy2}\\\nonumber
&+& \frac{\partial}{\partial X} \left( \begin{array}{c }
D_A\frac{\partial}{\partial X}P^{(0)}_A	 \\
D_B\frac{\partial}{\partial X}P^{(0)}_B
\end{array}
\right)
+  \frac{\partial}{\partial \tilde{x}} \left( \begin{array}{c }
D_A\frac{\partial}{\partial X}P^{(1)}_A	 \\
D_B\frac{\partial}{\partial X}P^{(1)}_B
\end{array}
\right)
+  \frac{\partial}{\partial X} \left( \begin{array}{c }
D_A\frac{\partial}{\partial \tilde{x}}P^{(1)}_A	 \\
D_B\frac{\partial}{\partial \tilde{x}}P^{(1)}_B
\end{array}
\right)
\, ,
\end{eqnarray}
\end{subequations}
respectively.
In \eqref{eq:hierarchy}, we introduced the 
tilde over the operator $\tilde{\OP{L}}^\dagger$  to
indicate that it acts on the small-scale variables
$\tilde{x}$.
Note that \eqref{eq:hierarchy0} is  the same
equation as \eqref{eq:fp_mat}, with the essential difference
that $P^{(0)}$ obeys periodic boundary conditions in the
variables $\tilde{ x }$.
Since we are interested in long-time solutions of \eqref{eq:hierarchy}
we consider the stationary state of the short-time processes, so that $\partial \boldsymbol{P}^{(a)}/\partial\theta = 0$ for all $a$ and there is no dependency on the fast time scale $\theta$.

\paragraph*{Order $\epsilon^0$.}
The stationary periodic solution in $\tilde{x}$  is a constant in $\tilde{x}$:  $\rho(\vartheta,\tau,X)$.
The chemical dynamics also relax to the steady state 
\begin{align}\label{eq:qAB}
    q_A(x,t)=&\frac{K_-}{K_++K_-}=\left(1+\frac{\phi_B}{\phi_A} \frac{r_+}{r_-}\right)^{-1}\,,\\ \nonumber
    q_B(x,t)=&
    \frac{K_+}{K_++K_-}=\left(1+\frac{\phi_A}{\phi_B} \frac{r_-}{r_+}\right)^{-1}
    \,,
\end{align}
and we have
\begin{equation}
\label{eq:zeroth-order}
P^{(0)}=\VEC{q}\rho(\vartheta,\tau,X)\,.
\end{equation}

\paragraph*{Order $\epsilon^1$.}
After relaxation of the fastest time variable the LHS of equation \ref{eq:hierarchy1}
reads: $- \left( \tilde{\OP{L}}^{\dagger} +  \TENSOR{K}  \right) P^{(1)}$.
According to the Fredholm alternative, a solution to the equation can exist only if the RHS is orthogonal to the left null space of the operator.
Such null space is spanned by constants multiplied by the 
 left null space of $ \TENSOR{K} $ which is given by $\hat{\VEC{q}}=(1,\;1)$.
 Taking the scalar product of $\hat{\VEC{q}}$ and the RHS of equation \ref{eq:hierarchy1} corresponds to summing its rows.
This solvability condition then requires
 \begin{equation}\label{eq:solv1}
\frac{\partial \rho}{\partial\vartheta}
+ \frac{\partial }{\partial X} \left(\langle v\rangle \rho\right)=0
\end{equation}
where 
\begin{equation}\label{eq:vbar}
\langle v\rangle=q_A v_A + q_B v_B\,,
\end{equation}
 which we obtained by exploiting that 
$P^{(0)}$ is independent of $\tilde{x}$.
Inserting Eq.~\eqref{eq:solv1} into Eq.~(\ref{eq:hierarchy1}), we obtain
\begin{eqnarray} \label{eq:second-order-equation}
- \left( \tilde{\OP{L}}^{\dagger} +  \TENSOR{K}  \right) P^{(1)}  
&=&
% \frac{\partial}{\partial X} \left(F\left( \begin{array}{c }
%(\mu_{+}-\bar{\mu})w_{+}	 \\
%(\mu_{-}-\bar{\mu}) w_{-}	\end{array}
%\right)\rho\right)\\\nonumber
\left( \begin{array}{c }
q_A	 \\
q_B		\end{array}
\right)\frac{\partial \langle v\rangle\rho }{\partial X}
- \frac{\partial}{\partial X} \left( \begin{array}{c }
v_Aq_A	 \\
 v_Bq_B	\end{array}
 \right)
 \rho\,.%\nonumber\\ \nonumber
% &=&
%  \frac{\partial}{\partial X} \left(F\left(-\frac{K_+}{K_++K_-}\right)\left(\mu_+-\mu_-\right)\VEC{m}_1\rho\right)
\end{eqnarray}
%  where, in moving to the last line we have exploited that the parameters related to the jump process ($K_-$, $K_+$, $w_-$, $w_+$, $\mu_-$, $\mu_+$,
% \ldots)
% are independent of space and
% where 
%  \begin{equation}
%   \VEC{m}_1=\left( \begin{array}{c }
% -1	 \\
%  1
% \end{array}
% \right)\frac{K_-}{K_-+K_+}
%   \end{equation}
%  is the right eigenvector associated with the non zero eigenvalue of $\TENSOR{K}$, which as required by the solvability condition, is orthogonal to $\hat{\VEC{w}}$ (see Appendix~\ref{sec:eigen}). 
Since the RHS does not depend on the small spatial scale $\tilde{x}$ we make the ansatz that also $\boldsymbol{P}^{(1)}$ is independent of them. 
To find $\VEC{P}^{(1)}$ we need to multiply the RHS by  the Green's function of the operator $\TENSOR{K}$
\begin{align}
\TENSOR{G} = \frac{1}{\left(K_++K_-\right)^2}\TENSOR{K}\,,
%  \left( \begin{array}{c c }
% K_+& -K_-	 \\
% -K_+& K_-	
% \end{array}
% \right)\
\end{align}
which is defined so that $\TENSOR{K}\TENSOR{G}=-\delta_{\alpha}^j+\VEC{q}\hat{\VEC{q}}$. 
 We obtain
\begin{align}\label{eq:p1}
\VEC{P}^{(1)}=\frac{\frac{\partial}{\partial X}\left(q_Aq_B(v_A-v_B)\rho\right)
+\frac{\partial q_A}{\partial X}\langle v\rangle\rho}
{K_-+K_+}
\left( \begin{array}{c }
-1 \\
1	\end{array}
\right)\,.
% )\\\nonumber
% &=&
% \frac{\partial\rho}{\partial X}F\left(\frac{K_+}{K_++K_-}\right)\left(\mu_+-\mu_-\right)
% %\frac{K_+}{\left(K_++K_-\right)^2}\left(\mu_+-\mu_-\right)
% \TENSOR{G}\VEC{m}_1\\\nonumber
% &=&\frac{\partial\rho}{\partial X}F\frac{K_+}{\left(K_++K_-\right)^2}\left(\mu_+-\mu_-\right)\VEC{m}_1
\end{align}

\paragraph*{Order $\epsilon^2$.}
%\textbf{HERE}
For equation (\ref{eq:hierarchy2}) to admit a solution, we need to require its RHS to be orthogonal to the left null space of $ \TENSOR{K} $ spanned by $\hat{\VEC{q}}$, which again, corresponds to summing over the rows of the RHS of (\ref{eq:hierarchy2}).
Plugging the solution for $\VEC{P}^{(1)}$~\eqref{eq:p1} and the condition~\eqref{eq:solv1})
%into eq. (\ref{eq:hierarchy2}) and imposing 
into 
the solvability condition yields:
\begin{align}\label{eq:solv2}
 \frac{\partial \rho}{\partial\tau} = -\frac{\partial}{\partial X}\left(v_A-v_B\right)P_A^{(1)}
%\left( \begin{array}{c }
%\mu_{+}P^{(1)}_+(x)	 \\
%\mu_{-} P^{(1)}_-(x)
%\end{array}
%\right)
+ \frac{\partial}{\partial X}\frac{\partial}{\partial X} \langle D\rangle\rho\,,
\end{align}
where
\begin{align}
\langle D\rangle =D_Aq_A+D_Bq_B\,.
\end{align}
%\begin{eqnarray}\label{eq:solv2}
% \frac{\partial \rho}{\partial\tau} &=& -F\frac{\partial}{\partial X}\left(\mu_+-\overline{\mu},\, \mu_--\overline{\mu}  \right)P^{(1)}
%%\left( \begin{array}{c }
%%\mu_{+}P^{(1)}_+(x)	 \\
%%\mu_{-} P^{(1)}_-(x)
%%\end{array}
%%\right)
%\\\nonumber
%&+&T \frac{\partial}{\partial X}\frac{\partial}{\partial X} \left(\mu_+,\, \mu_- \right)P^{(0)}\\\nonumber
% &=&
%  \frac{\partial^2\rho}{\partial X^2}\left(T\overline{\mu}+F^2\frac{K_+K_-}{\left(K_++K_-\right)^3}\left(\mu_+-\mu_-\right)^2 \right)
%%\, ,
%\end{eqnarray}
%where we have made use of the fact that $\left(\mu_+-\overline{\mu},\, \mu_--\overline{\mu}  \right)=-(\mu_+-\mu_-)\frac{K_-}{K_++K_-}\hat{\VEC{n}}_1$.
Inserting the explicit expression of $P_A^{(1)}$, we obtain 
\begin{align}\label{eq:solv2_exp}
 \frac{\partial \rho}{\partial\tau} = -\frac{\partial}{\partial X}\left(v_s+\frac{\partial D^{\mathrm{eff}}}{\partial X}\right)\rho%\left( \begin{array}{c }
%\mu_{+}P^{(1)}_+(x)	 \\
%\mu_{-} P^{(1)}_-(x)
%\end{array}
%\right)
+ \frac{\partial}{\partial X}\frac{\partial}{\partial X} D^{\mathrm{eff}} \rho\,,
\end{align}
where $D^{\mathrm{eff}} = \langle D\rangle+D^{\mathrm{enh}}$ and
\begin{align}\label{eq:Denh_app}
D^{\mathrm{enh}} = \frac{ q_Aq_B}{K_++K_-}\left(v_A-v_B\right)^2
\end{align}
is the effective enhanced diffusion coefficient.
This enhancement is predicted to be small
 because the single-molecule drifts appearing in Eq.~\eqref{eq:Denh_app} are closely related to the spatial dependency of the reaction rates. Indeed, from Eq.~\eqref{eq:vandK},
 %
 % combining the local detailed balance condition, Eq.~\eqref{eq:detailed_balance}, with the definition of  the single-molecule reaction rates, Eq.~\eqref{eq:K}, reveals that  the difference in the drifts is related to the gradients of the reaction rates
 %
 % computing the gradients of the single-molecule reaction rates given in Eq.\eqref{eq:K_explicit}, and recalling the expression for the drifts \eqref{eq:vA-vB_mu}, we see that 
$v_A-v_B=D\nabla\ln K_-/K_+$ so that the diffusion enhancement is
\begin{align}\label{eq:Denh_ratio_new}
%\frac{ q_Aq_B}{D\left(K_++K_-\right)}\left(v_A-v_B\right)^2
D^{\mathrm{enh}}=q_Aq_B\left(\ell_D\nabla\ln\frac{K_-}{K_+}\right)^2\,,
\end{align}
which is small whenever the condition for the validity of the multiple-scale approach
$K/\nabla K\gg \ell_D$ is satisfied.

%given in Eq.~\eqref{eq:eff_D} in the main text and 
The spurious drift is
%\begin{equation}\label{eq:lang_eff}
%d x^{eff}=F\bar{\mu}dt+\underbrace{\sqrt{2T\bar{\mu}+2F^2(\mu_{+}-\mu_{-})^2\frac{K_-K_+}{(K_-+K_+)^3}}}_{\sqrt{2D_{eff}}}dW'_t
%\end{equation}
%showing that the effective equation evolves with average
%\begin{equation}\label{eq:ave_m}
%\langle \Delta x^{eff}\rangle=\bar{\mu}Ft
%\end{equation}
%and variance
%\begin{eqnarray}\label{eq:var_m}
%&&\left\langle\left( \Delta x^{eff}-\langle \Delta x^{eff}\rangle\right)^2\right\rangle=\\\nonumber
%&&2\left[T\bar{\mu}+F^2(\mu_{+}-\mu_{-})^2\frac{K_-K_+}{(K_-+K_+)^3}\right]t
%\end{eqnarray}
% \begin{align}
%     {v}_s=\left({v}_A-{v}_B\right) \left\{\nabla\left[\frac{q_Aq_B}{K_++K_-}\left({v}_A-{v}_B\right)\right]+\frac{q_B{v}_A+q_A{v}_B}{K_++K_-}\nabla q_B\right\}\,.
%     % \\
%     %     \VEC{v}_s=\left(\VEC{v}_A-\VEC{v}_B\right)\sum_\alpha \left\{\frac{\partial}{\partial \phi_\alpha}\left[\frac{wq_A}{K_++K_-}\left(\VEC{v}_A-\VEC{v}_B\right)\right]+\frac{w\VEC{v}_A+q_A\VEC{v}_B}{K_++K_-}\frac{\partial w}{\partial \phi_\alpha}\right\}\nabla \phi_\alpha
% \end{align}
% which can also be written as
\begin{align}\label{eq:eff_drift_s}
    {v}_s=%\nabla\left[\frac{q_Aq_B}{K_++K_-}\left({v}_A-{v}_B\right)^2\right]+
    \frac{{v}_A-{v}_B}{K_++K_-} \left[-q_Aq_B\nabla\left({v}_A-{v}_B\right)+(q_B{v}_A+q_A{v}_B)\nabla q_B\right]-\nabla \langle D\rangle\,.
\end{align}
Summing the two solvability conditions \eqref{eq:solv1} and \eqref{eq:solv2} and returning to the original variables gives the Fokker-Planck equation corresponding to the effective Langevin equation given in Eq.~\eqref{eq:eff_lang} in the main text.
\subsection{Multi-dimensional case}\label{app:multi_multi}
For a spherical object, the diffusion coefficient is proportional to the identity.
The different spatial directions are then decoupled and to discuss a multidimensional scenario, we just have to repeat the derivation above for the different dimension.
The resulting effective equation may feature anisotropic diffusion, in case the drifts have different magnitudes in different directions.
In components, the effective Langevin equation reads
\begin{align}
    \frac{dx_i}{dt}=v^{\mathrm{eff}}_i+\frac{\partial }{\partial x_j}D^{\mathrm{eff}}_{ij}+\beta^{\mathrm{eff}}_{ij}w_j
\end{align}
where $\beta^{\mathrm{eff}}_{ik}\beta^{\mathrm{eff}}_{jk}=2D^{\mathrm{eff}}_{ij}$.
The effective drift reads
\begin{align}
    \VEC{v}^{\mathrm{eff}}=\langle \VEC{v} \rangle+ \VEC{v}_s\,,
\end{align}
with
\begin{align}\label{eq:eff_drift_s_multid}
\langle \VEC{v} \rangle&=q_A\VEC{v}_A+q_B\VEC{v}_B\\
    \VEC{v}_s&=\frac{\VEC{v}_A-\VEC{v}_B}{K_++K_-} \left[-q_Aq_B\nabla\left(\VEC{v}_A-\VEC{v}_B\right)+(q_B\VEC{v}_A+q_A\VEC{v}_B)\nabla q_B\right]-
    \frac{\partial }{\partial x_j}<D>_{ij}\nabla \langle D\rangle\,\,.
    % \\
    %     \VEC{v}_s=\left(\VEC{v}_A-\VEC{v}_B\right)\sum_\alpha \left\{\frac{\partial}{\partial \phi_\alpha}\left[\frac{wq_A}{K_++K_-}\left(\VEC{v}_A-\VEC{v}_B\right)\right]+\frac{w\VEC{v}_A+q_A\VEC{v}_B}{K_++K_-}\frac{\partial w}{\partial \phi_\alpha}\right\}\nabla \phi_\alpha
\end{align}
The effective diffusion matrix is diagonal with entries $D^{\mathrm{eff}}_i$, where
\begin{align}
\label{eq:eff_D_multi}
D^{\mathrm{eff}}_i &= \langle D_i\rangle + \frac{ q_Aq_B}{K_++K_-}\left(v_{A,\,i}-v_{B,\,i}\right)^2\,,
% \\\nonumber
%                  &=\langle D\rangle + \frac{ q_Aq_B}{K_++K_-}\left(m\nabla\lambda_\chi\right)^2\,,
\end{align}
where $v_{A,\,i}$ ($v_{B,\,i}$) is the $i$-th spatial component of the drift acting on molecule $A$ ($B$).
\section{Numerical study of a ternary mixture with a chemical reaction}\label{app:numerics}
\subsection{Volume fraction dynamics}
In order to solve the Cahn-Hilliard problem for the volume-fraction profiles, we need to solve two fourth-order PDEs, one for each species A and B [see Eq.~\eqref{eq:cont_AB} and \eqref{eq:j}]. We use a finite difference scheme, using central differences and the forward Euler method in time. The full code can be found at \cite{Hubatsch2025Https://git.mpi-cbg.de/hubatsch/chemical_reactions} %\href{fhttps://git.mpi-cbg.de/hubatsch/chemical_reactions}{online} 
(tested with Julia 1.11).
\subsubsection{Equilibrium case}\label{app:num_eq}
We use no-flux boundary conditions and set the second derivatives to zero. Initial conditions are opposite ramps of $\phi_A$ and $\phi_B$, with an average volume fraction of 0.042 and 0.45, respectively, inspired by \cite{Bauermann2022b}. The parameters used are as follows: 
$\omega_A=\omega_B=\unit[0.5]{k_BT}$,
$\kappa_{A}=\kappa_{B}=\unit[0.003]{\mu m^{-2}}$,
$\chi_{A0}=-\unit[1]{k_BT\mu m^{-3}}$, $\chi_{B0}=\unit[3]{k_BT\mu m^{-3}}$, $\chi_{AB}=0$, 
$k=\unit[10]{s^{-1}}$, 
$m_{A}=m_{B}=m_{AB}=\unit[0.1]{\mu m^2s^{-1}(k_BT)^{-1}}$, system size is 
$L=\unit[1]{\mu m}$, %and we have $k_{B}T=1$ 
and $\nu=\nu_0=\unit[1]{\mu m^3}$.
\subsubsection{Nonequilibrium enzyme}
% $\chi=\unit[6.5]{k_BT\mu m^{-3}}$, $\omega_A=\omega_B=\unit[0.5]{k_BT}$ and $\kappa_A=\kappa_B=\unit[2k_BT]{\mu m^{-2}}$
%
We enforce Dirichlet boundary conditions for the volume fractions 
$\phi_{A}(0) = 0.68$, $\phi_{A}(1) = 0.031$, $\phi_{B}(0) = 0.31$, $\phi_{B}(1) = 0.036$, and set the second derivatives of the volume fraction to zero at the boundary. We use the following parameters: $\omega_A=\omega_B=\unit[0.5]{k_BT}$, $\kappa_{A}=\kappa_{B}=\unit[0.15]{\mu m^{-2}}$, $\chi_{A0}=\unit[1]{k_BT\mu m^{-3}}$, $\chi_{B0}=\unit[3]{k_BT\mu m^{-3}}$, $\chi_{AB}=-\unit[3]{k_BT\mu m^{-3}}$, 
$k=\unit[25]{s^{-1}}$,
, $m_{A}=m_{B}=m_{AB}=0.1$, system size is $L=\unit[1]{\mu m}$, %and we have $k_{B}T=1$ 
and $\nu=\nu_0=\unit[1]{\mu m^3}$.
\subsection{Stochastic simulations}
We performed stochastic simulations using an Euler-Maruyama algorithm~\cite{Kloeden1992} with timestep $\delta t=\unit[10^{-5}]{s}$ and reflecting boundary conditions set at $x_L=\unit[0.01]{\mu m}$ and $x_R=\unit[0.99]{\mu m}$.
For the equilibrium simulation reported in fig.~\ref{fig:eq} we simulated $4\times 10^4$ stochastic trajectories, each for a time $T=\unit[30]{s}$.\\
The non-equilibrium scenario reported in Fig.~\ref{fig:neq} is obtained from the simulation of $10^7$ stochastic trajectories, each for a time $T=\unit[0.1]{s}$. 
\subsection{Drift and diffusion measurements from stochastic trajectories}
The empirical drift is computed as a local mean velocity for a fixed starting position and chemical state of the molecule
\begin{align}\label{eq:mean}
w_{\Delta t}(x_\alpha)=
      \left\langle x(t+\Delta t)-x(t)|x(t)=x_\alpha\right\rangle/\Delta t\,,
\end{align}
where $\langle\ldots\rangle$ is an ensemble average, which we perform over repeated simulations and we condition on the starting point of the trajectory increase $x(t)$ to be equal to $x_\alpha$.
The drift, irrespective of the chemical state, is 
\begin{align}\label{eq:mean_irr}
w_{\Delta t}(x)=
      \left\langle x(t+\Delta t)-x(t)|x(t)=x\right\rangle/\Delta t\,.
\end{align}
Note that these drift estimations return the drifts in the Ito discretization because the averages are conditioned on the starting position $x(t)$. They, therefore, estimate $v+\nabla D$.
If we measure the system on sufficiently long scales, we can estimate the effective drift from Eq.~\eqref{eq:mean_irr}, which will return an estimate of
\begin{align}\label{eq:vtilde}
    \tilde{v}\equiv v^\mathrm{eff}+\nabla D^\mathrm{eff}\,.
\end{align}

The empirical diffusion coefficient, conditioned on the chemical state of the molecule, is computed as
\begin{align}\label{eq:MSD}
   \mathcal{D}_{\Delta t}(x_\alpha)=\left\langle \left[x(t+\Delta t)-x(t)-w_{\Delta t}(x_\alpha)\Delta t\right]^2|x(t)=x_\alpha\right\rangle/2\Delta t\,,
\end{align}
and the one irrespective of the chemical state as
\begin{align}\label{eq:MSD_irr}
   \mathcal{D}_{\Delta t}(x)=\left\langle \left[x(t+\Delta t)-x(t)-w_{\Delta t}(x)\Delta t\right]^2|x(t)=x\right\rangle/2\Delta t\,.
\end{align}
To observe the effective dynamics, we need to explore longer time scales. Measuring a diffusion coefficient with a finite time resolution can prove difficult, and it is necessary to take into account a correction to the estimator of Eq.~\ref{eq:MSD}. The corrected estimator was proposed in Ref.~\cite{Ragwitz2001IndispensableData} and reads
\begin{align}\label{eq:effD_est}
\mathcal{D}_c(x)=\frac{\mathcal{D}_{\Delta t}}{1+ \frac{dw(x)}{dx}\Delta t} \,,
\end{align}
where $dw(x)/dx$ is the derivative of the estimated drift, which we compute empirically.
\bibliography{references}

\begin{thebibliography}{61}%
\makeatletter
\providecommand \@ifxundefined [1]{%
 \@ifx{#1\undefined}
}%
\providecommand \@ifnum [1]{%
 \ifnum #1\expandafter \@firstoftwo
 \else \expandafter \@secondoftwo
 \fi
}%
\providecommand \@ifx [1]{%
 \ifx #1\expandafter \@firstoftwo
 \else \expandafter \@secondoftwo
 \fi
}%
\providecommand \natexlab [1]{#1}%
\providecommand \enquote  [1]{``#1''}%
\providecommand \bibnamefont  [1]{#1}%
\providecommand \bibfnamefont [1]{#1}%
\providecommand \citenamefont [1]{#1}%
\providecommand \href@noop [0]{\@secondoftwo}%
\providecommand \href [0]{\begingroup \@sanitize@url \@href}%
\providecommand \@href[1]{\@@startlink{#1}\@@href}%
\providecommand \@@href[1]{\endgroup#1\@@endlink}%
\providecommand \@sanitize@url [0]{\catcode `\\12\catcode `\$12\catcode
  `\&12\catcode `\#12\catcode `\^12\catcode `\_12\catcode `\%12\relax}%
\providecommand \@@startlink[1]{}%
\providecommand \@@endlink[0]{}%
\providecommand \url  [0]{\begingroup\@sanitize@url \@url }%
\providecommand \@url [1]{\endgroup\@href {#1}{\urlprefix }}%
\providecommand \urlprefix  [0]{URL }%
\providecommand \Eprint [0]{\href }%
\providecommand \doibase [0]{https://doi.org/}%
\providecommand \selectlanguage [0]{\@gobble}%
\providecommand \bibinfo  [0]{\@secondoftwo}%
\providecommand \bibfield  [0]{\@secondoftwo}%
\providecommand \translation [1]{[#1]}%
\providecommand \BibitemOpen [0]{}%
\providecommand \bibitemStop [0]{}%
\providecommand \bibitemNoStop [0]{.\EOS\space}%
\providecommand \EOS [0]{\spacefactor3000\relax}%
\providecommand \BibitemShut  [1]{\csname bibitem#1\endcsname}%
\let\auto@bib@innerbib\@empty
%</preamble>
\bibitem [{\citenamefont {Banani}\ \emph {et~al.}(2017)\citenamefont {Banani},
  \citenamefont {Lee}, \citenamefont {Hyman},\ and\ \citenamefont
  {Rosen}}]{Banani2017}%
  \BibitemOpen
  \bibfield  {author} {\bibinfo {author} {\bibfnamefont {S.~F.}\ \bibnamefont
  {Banani}}, \bibinfo {author} {\bibfnamefont {H.~O.}\ \bibnamefont {Lee}},
  \bibinfo {author} {\bibfnamefont {A.~A.}\ \bibnamefont {Hyman}},\ and\
  \bibinfo {author} {\bibfnamefont {M.~K.}\ \bibnamefont {Rosen}},\ }\bibfield
  {title} {\bibinfo {title} {{Biomolecular condensates: Organizers of cellular
  biochemistry}},\ }\href {https://doi.org/10.1038/nrm.2017.7} {\bibfield
  {journal} {\bibinfo  {journal} {Nature Reviews Molecular Cell Biology}\
  }\textbf {\bibinfo {volume} {18}},\ \bibinfo {pages} {285} (\bibinfo {year}
  {2017})}\BibitemShut {NoStop}%
\bibitem [{\citenamefont {Alberti}\ \emph {et~al.}(2019)\citenamefont
  {Alberti}, \citenamefont {Gladfelter},\ and\ \citenamefont
  {Mittag}}]{Alberti2019}%
  \BibitemOpen
  \bibfield  {author} {\bibinfo {author} {\bibfnamefont {S.}~\bibnamefont
  {Alberti}}, \bibinfo {author} {\bibfnamefont {A.}~\bibnamefont
  {Gladfelter}},\ and\ \bibinfo {author} {\bibfnamefont {T.}~\bibnamefont
  {Mittag}},\ }\bibfield  {title} {\bibinfo {title} {{Considerations and
  Challenges in Studying Liquid-Liquid Phase Separation and Biomolecular
  Condensates}},\ }\href {https://doi.org/10.1016/j.cell.2018.12.035}
  {\bibfield  {journal} {\bibinfo  {journal} {Cell}\ }\textbf {\bibinfo
  {volume} {176}},\ \bibinfo {pages} {419} (\bibinfo {year}
  {2019})}\BibitemShut {NoStop}%
\bibitem [{\citenamefont {Patel}\ \emph {et~al.}(2015)\citenamefont {Patel},
  \citenamefont {Lee}, \citenamefont {Jawerth}, \citenamefont {Maharana},
  \citenamefont {Jahnel}, \citenamefont {Hein}, \citenamefont {Stoynov},
  \citenamefont {Mahamid}, \citenamefont {Saha}, \citenamefont {Franzmann},
  \citenamefont {Pozniakovski}, \citenamefont {Poser}, \citenamefont
  {Maghelli}, \citenamefont {Royer}, \citenamefont {Weigert}, \citenamefont
  {Myers}, \citenamefont {Grill}, \citenamefont {Drechsel}, \citenamefont
  {Hyman},\ and\ \citenamefont {Alberti}}]{Patel2015AMutation}%
  \BibitemOpen
  \bibfield  {author} {\bibinfo {author} {\bibfnamefont {A.}~\bibnamefont
  {Patel}}, \bibinfo {author} {\bibfnamefont {H.~O.}\ \bibnamefont {Lee}},
  \bibinfo {author} {\bibfnamefont {L.}~\bibnamefont {Jawerth}}, \bibinfo
  {author} {\bibfnamefont {S.}~\bibnamefont {Maharana}}, \bibinfo {author}
  {\bibfnamefont {M.}~\bibnamefont {Jahnel}}, \bibinfo {author} {\bibfnamefont
  {M.~Y.}\ \bibnamefont {Hein}}, \bibinfo {author} {\bibfnamefont
  {S.}~\bibnamefont {Stoynov}}, \bibinfo {author} {\bibfnamefont
  {J.}~\bibnamefont {Mahamid}}, \bibinfo {author} {\bibfnamefont
  {S.}~\bibnamefont {Saha}}, \bibinfo {author} {\bibfnamefont {T.~M.}\
  \bibnamefont {Franzmann}}, \bibinfo {author} {\bibfnamefont {A.}~\bibnamefont
  {Pozniakovski}}, \bibinfo {author} {\bibfnamefont {I.}~\bibnamefont {Poser}},
  \bibinfo {author} {\bibfnamefont {N.}~\bibnamefont {Maghelli}}, \bibinfo
  {author} {\bibfnamefont {L.~A.}\ \bibnamefont {Royer}}, \bibinfo {author}
  {\bibfnamefont {M.}~\bibnamefont {Weigert}}, \bibinfo {author} {\bibfnamefont
  {E.~W.}\ \bibnamefont {Myers}}, \bibinfo {author} {\bibfnamefont
  {S.}~\bibnamefont {Grill}}, \bibinfo {author} {\bibfnamefont
  {D.}~\bibnamefont {Drechsel}}, \bibinfo {author} {\bibfnamefont {A.~A.}\
  \bibnamefont {Hyman}},\ and\ \bibinfo {author} {\bibfnamefont
  {S.}~\bibnamefont {Alberti}},\ }\bibfield  {title} {\bibinfo {title} {{A
  Liquid-to-Solid Phase Transition of the ALS Protein FUS Accelerated by
  Disease Mutation}},\ }\href {https://doi.org/10.1016/j.cell.2015.07.047}
  {\bibfield  {journal} {\bibinfo  {journal} {Cell}\ }\textbf {\bibinfo
  {volume} {162}},\ \bibinfo {pages} {1066} (\bibinfo {year}
  {2015})}\BibitemShut {NoStop}%
\bibitem [{\citenamefont {Alberti}\ and\ \citenamefont
  {Hyman}(2021)}]{Alberti2021BiomolecularAgeing}%
  \BibitemOpen
  \bibfield  {author} {\bibinfo {author} {\bibfnamefont {S.}~\bibnamefont
  {Alberti}}\ and\ \bibinfo {author} {\bibfnamefont {A.~A.}\ \bibnamefont
  {Hyman}},\ }\href {https://doi.org/10.1038/s41580-020-00326-6} {\bibinfo
  {title} {{Biomolecular condensates at the nexus of cellular stress, protein
  aggregation disease and ageing}}} (\bibinfo {year} {2021})\BibitemShut
  {NoStop}%
\bibitem [{\citenamefont {Tsang}\ \emph {et~al.}(2020)\citenamefont {Tsang},
  \citenamefont {Priti{\v{s}}anac}, \citenamefont {Scherer}, \citenamefont
  {Moses},\ and\ \citenamefont {Forman-Kay}}]{Tsang2020PhaseMutations}%
  \BibitemOpen
  \bibfield  {author} {\bibinfo {author} {\bibfnamefont {B.}~\bibnamefont
  {Tsang}}, \bibinfo {author} {\bibfnamefont {I.}~\bibnamefont
  {Priti{\v{s}}anac}}, \bibinfo {author} {\bibfnamefont {S.~W.}\ \bibnamefont
  {Scherer}}, \bibinfo {author} {\bibfnamefont {A.~M.}\ \bibnamefont {Moses}},\
  and\ \bibinfo {author} {\bibfnamefont {J.~D.}\ \bibnamefont {Forman-Kay}},\
  }\href {https://doi.org/10.1016/j.cell.2020.11.050} {\bibinfo {title} {{Phase
  Separation as a Missing Mechanism for Interpretation of Disease Mutations}}}
  (\bibinfo {year} {2020})\BibitemShut {NoStop}%
\bibitem [{\citenamefont {Hyman}\ \emph {et~al.}(2014)\citenamefont {Hyman},
  \citenamefont {Weber},\ and\ \citenamefont {J{\"{u}}licher}}]{Hyman2014}%
  \BibitemOpen
  \bibfield  {author} {\bibinfo {author} {\bibfnamefont {A.~A.}\ \bibnamefont
  {Hyman}}, \bibinfo {author} {\bibfnamefont {C.~A.}\ \bibnamefont {Weber}},\
  and\ \bibinfo {author} {\bibfnamefont {F.}~\bibnamefont {J{\"{u}}licher}},\
  }\bibfield  {title} {\bibinfo {title} {{Liquid-Liquid Phase Separation in
  Biology}},\ }\href {https://doi.org/10.1146/annurev-cellbio-100913-013325}
  {\bibfield  {journal} {\bibinfo  {journal} {Annual Review of Cell and
  Developmental Biology}\ }\textbf {\bibinfo {volume} {30}},\ \bibinfo {pages}
  {39} (\bibinfo {year} {2014})}\BibitemShut {NoStop}%
\bibitem [{\citenamefont {Weber}\ \emph {et~al.}(2019)\citenamefont {Weber},
  \citenamefont {Zwicker}, \citenamefont {J{\"{u}}licher},\ and\ \citenamefont
  {Lee}}]{Weber2019}%
  \BibitemOpen
  \bibfield  {author} {\bibinfo {author} {\bibfnamefont {C.~A.}\ \bibnamefont
  {Weber}}, \bibinfo {author} {\bibfnamefont {D.}~\bibnamefont {Zwicker}},
  \bibinfo {author} {\bibfnamefont {F.}~\bibnamefont {J{\"{u}}licher}},\ and\
  \bibinfo {author} {\bibfnamefont {C.~F.}\ \bibnamefont {Lee}},\ }\bibfield
  {title} {\bibinfo {title} {{Physics of active emulsions}},\ }\href
  {https://doi.org/10.1088/1361-6633/ab052b} {\bibfield  {journal} {\bibinfo
  {journal} {Reports on Progress in Physics}\ }\textbf {\bibinfo {volume}
  {82}},\ \bibinfo {pages} {064601} (\bibinfo {year} {2019})}\BibitemShut
  {NoStop}%
\bibitem [{\citenamefont {Fritsch}\ \emph {et~al.}(2021)\citenamefont
  {Fritsch}, \citenamefont {Diaz-Delgadillo}, \citenamefont {Adame-Arana},
  \citenamefont {Hoege}, \citenamefont {Mittasch}, \citenamefont {Kreysing},
  \citenamefont {Leaver}, \citenamefont {Hyman}, \citenamefont
  {J{\"{u}}licher},\ and\ \citenamefont {Weber}}]{Fritsch2021}%
  \BibitemOpen
  \bibfield  {author} {\bibinfo {author} {\bibfnamefont {A.~W.}\ \bibnamefont
  {Fritsch}}, \bibinfo {author} {\bibfnamefont {A.~F.}\ \bibnamefont
  {Diaz-Delgadillo}}, \bibinfo {author} {\bibfnamefont {O.}~\bibnamefont
  {Adame-Arana}}, \bibinfo {author} {\bibfnamefont {C.}~\bibnamefont {Hoege}},
  \bibinfo {author} {\bibfnamefont {M.}~\bibnamefont {Mittasch}}, \bibinfo
  {author} {\bibfnamefont {M.}~\bibnamefont {Kreysing}}, \bibinfo {author}
  {\bibfnamefont {M.}~\bibnamefont {Leaver}}, \bibinfo {author} {\bibfnamefont
  {A.~A.}\ \bibnamefont {Hyman}}, \bibinfo {author} {\bibfnamefont
  {F.}~\bibnamefont {J{\"{u}}licher}},\ and\ \bibinfo {author} {\bibfnamefont
  {C.~A.}\ \bibnamefont {Weber}},\ }\bibfield  {title} {\bibinfo {title}
  {{Local thermodynamics govern formation and dissolution of Caenorhabditis
  elegans P granule condensates}},\ }\href
  {https://doi.org/10.1073/PNAS.2102772118} {\bibfield  {journal} {\bibinfo
  {journal} {Proceedings of the National Academy of Sciences}\ }\textbf
  {\bibinfo {volume} {118}},\ \bibinfo {pages} {e2102772118} (\bibinfo {year}
  {2021})}\BibitemShut {NoStop}%
\bibitem [{\citenamefont {J{\"{u}}licher}\ and\ \citenamefont
  {Weber}(2024)}]{Julicher2024DropletSeparation}%
  \BibitemOpen
  \bibfield  {author} {\bibinfo {author} {\bibfnamefont {F.}~\bibnamefont
  {J{\"{u}}licher}}\ and\ \bibinfo {author} {\bibfnamefont {C.~A.}\
  \bibnamefont {Weber}},\ }\href
  {https://doi.org/10.1146/annurev-conmatphys-031720-032917} {\bibinfo {title}
  {{Droplet Physics and Intracellular Phase Separation}}} (\bibinfo {year}
  {2024})\BibitemShut {NoStop}%
\bibitem [{\citenamefont {Hubatsch}\ \emph {et~al.}(2021)\citenamefont
  {Hubatsch}, \citenamefont {Jawerth}, \citenamefont {Love}, \citenamefont
  {Bauermann}, \citenamefont {Tang}, \citenamefont {Bo}, \citenamefont
  {Hyman},\ and\ \citenamefont {Weber}}]{Hubatsch2021QuantitativeCondensates}%
  \BibitemOpen
  \bibfield  {author} {\bibinfo {author} {\bibfnamefont {L.}~\bibnamefont
  {Hubatsch}}, \bibinfo {author} {\bibfnamefont {L.~M.}\ \bibnamefont
  {Jawerth}}, \bibinfo {author} {\bibfnamefont {C.}~\bibnamefont {Love}},
  \bibinfo {author} {\bibfnamefont {J.}~\bibnamefont {Bauermann}}, \bibinfo
  {author} {\bibfnamefont {T.~Y.}\ \bibnamefont {Tang}}, \bibinfo {author}
  {\bibfnamefont {S.}~\bibnamefont {Bo}}, \bibinfo {author} {\bibfnamefont
  {A.~A.}\ \bibnamefont {Hyman}},\ and\ \bibinfo {author} {\bibfnamefont
  {C.~A.}\ \bibnamefont {Weber}},\ }\bibfield  {title} {\bibinfo {title}
  {{Quantitative theory for the diffusive dynamics of liquid condensates}},\
  }\bibfield  {journal} {\bibinfo  {journal} {eLife}\ }\textbf {\bibinfo
  {volume} {10}},\ \href {https://doi.org/10.7554/ELIFE.68620}
  {10.7554/ELIFE.68620} (\bibinfo {year} {2021})\BibitemShut {NoStop}%
\bibitem [{\citenamefont {Bauermann}\ \emph
  {et~al.}(2022{\natexlab{a}})\citenamefont {Bauermann}, \citenamefont {Laha},
  \citenamefont {McCall}, \citenamefont {J{\"{u}}licher},\ and\ \citenamefont
  {Weber}}]{Bauermann2022b}%
  \BibitemOpen
  \bibfield  {author} {\bibinfo {author} {\bibfnamefont {J.}~\bibnamefont
  {Bauermann}}, \bibinfo {author} {\bibfnamefont {S.}~\bibnamefont {Laha}},
  \bibinfo {author} {\bibfnamefont {P.~M.}\ \bibnamefont {McCall}}, \bibinfo
  {author} {\bibfnamefont {F.}~\bibnamefont {J{\"{u}}licher}},\ and\ \bibinfo
  {author} {\bibfnamefont {C.~A.}\ \bibnamefont {Weber}},\ }\bibfield  {title}
  {\bibinfo {title} {{Chemical Kinetics and Mass Action in Coexisting
  Phases}},\ }\bibfield  {journal} {\bibinfo  {journal} {Journal of the
  American Chemical Society}\ }\textbf {\bibinfo {volume} {144}},\ \href
  {https://doi.org/10.1021/jacs.2c06265} {10.1021/jacs.2c06265} (\bibinfo
  {year} {2022}{\natexlab{a}})\BibitemShut {NoStop}%
\bibitem [{\citenamefont {Harmon}\ and\ \citenamefont
  {J{\"{u}}licher}(2022)}]{Harmon2022MolecularDroplets}%
  \BibitemOpen
  \bibfield  {author} {\bibinfo {author} {\bibfnamefont {T.~S.}\ \bibnamefont
  {Harmon}}\ and\ \bibinfo {author} {\bibfnamefont {F.}~\bibnamefont
  {J{\"{u}}licher}},\ }\bibfield  {title} {\bibinfo {title} {{Molecular
  Assembly Lines in Active Droplets}},\ }\href
  {https://doi.org/10.1103/PHYSREVLETT.128.108102/FIGURES/3/MEDIUM} {\bibfield
  {journal} {\bibinfo  {journal} {Physical Review Letters}\ }\textbf {\bibinfo
  {volume} {128}},\ \bibinfo {pages} {108102} (\bibinfo {year}
  {2022})}\BibitemShut {NoStop}%
\bibitem [{\citenamefont {Nott}\ \emph {et~al.}(2016)\citenamefont {Nott},
  \citenamefont {Craggs},\ and\ \citenamefont
  {Baldwin}}]{Nott2016MembranelessFilters}%
  \BibitemOpen
  \bibfield  {author} {\bibinfo {author} {\bibfnamefont {T.~J.}\ \bibnamefont
  {Nott}}, \bibinfo {author} {\bibfnamefont {T.~D.}\ \bibnamefont {Craggs}},\
  and\ \bibinfo {author} {\bibfnamefont {A.~J.}\ \bibnamefont {Baldwin}},\
  }\bibfield  {title} {\bibinfo {title} {{Membraneless organelles can melt
  nucleic acid duplexes and act as biomolecular filters}},\ }\href
  {https://doi.org/10.1038/nchem.2519} {\bibfield  {journal} {\bibinfo
  {journal} {Nature Chemistry 2016 8:6}\ }\textbf {\bibinfo {volume} {8}},\
  \bibinfo {pages} {569} (\bibinfo {year} {2016})}\BibitemShut {NoStop}%
\bibitem [{\citenamefont {Shelest}\ \emph {et~al.}(2024)\citenamefont
  {Shelest}, \citenamefont {Roy}, \citenamefont {Busiello},\ and\ \citenamefont
  {Rios}}]{Shelest2024PhaseFluxes}%
  \BibitemOpen
  \bibfield  {author} {\bibinfo {author} {\bibfnamefont {A.}~\bibnamefont
  {Shelest}}, \bibinfo {author} {\bibfnamefont {H.~L.}\ \bibnamefont {Roy}},
  \bibinfo {author} {\bibfnamefont {D.~M.}\ \bibnamefont {Busiello}},\ and\
  \bibinfo {author} {\bibfnamefont {P.~D.~L.}\ \bibnamefont {Rios}},\
  }\bibfield  {title} {\bibinfo {title} {{Phase boundaries promote chemical
  reactions through localized fluxes}},\ }\href
  {https://arxiv.org/abs/2406.19266v2 http://arxiv.org/abs/2406.19266}
  {\bibfield  {journal} {\bibinfo  {journal} {arXiv}\ }\textbf {\bibinfo
  {volume} {2406.19266}} (\bibinfo {year} {2024})}\BibitemShut {NoStop}%
\bibitem [{\citenamefont {Hondele}\ \emph {et~al.}(2019)\citenamefont
  {Hondele}, \citenamefont {Sachdev}, \citenamefont {Heinrich}, \citenamefont
  {Wang}, \citenamefont {Vallotton}, \citenamefont {Fontoura},\ and\
  \citenamefont {Weis}}]{Hondele2019DEAD-boxOrganelles}%
  \BibitemOpen
  \bibfield  {author} {\bibinfo {author} {\bibfnamefont {M.}~\bibnamefont
  {Hondele}}, \bibinfo {author} {\bibfnamefont {R.}~\bibnamefont {Sachdev}},
  \bibinfo {author} {\bibfnamefont {S.}~\bibnamefont {Heinrich}}, \bibinfo
  {author} {\bibfnamefont {J.}~\bibnamefont {Wang}}, \bibinfo {author}
  {\bibfnamefont {P.}~\bibnamefont {Vallotton}}, \bibinfo {author}
  {\bibfnamefont {B.~M.}\ \bibnamefont {Fontoura}},\ and\ \bibinfo {author}
  {\bibfnamefont {K.}~\bibnamefont {Weis}},\ }\bibfield  {title} {\bibinfo
  {title} {{DEAD-box ATPases are global regulators of phase-separated
  organelles}},\ }\href {https://doi.org/10.1038/s41586-019-1502-y} {\bibfield
  {journal} {\bibinfo  {journal} {Nature}\ }\textbf {\bibinfo {volume} {573}},\
  \bibinfo {pages} {144} (\bibinfo {year} {2019})}\BibitemShut {NoStop}%
\bibitem [{\citenamefont {Cotton}\ \emph
  {et~al.}(2022{\natexlab{a}})\citenamefont {Cotton}, \citenamefont
  {Golestanian},\ and\ \citenamefont
  {Agudo-Canalejo}}]{Cotton2022Catalysis-InducedActivity}%
  \BibitemOpen
  \bibfield  {author} {\bibinfo {author} {\bibfnamefont {M.~W.}\ \bibnamefont
  {Cotton}}, \bibinfo {author} {\bibfnamefont {R.}~\bibnamefont
  {Golestanian}},\ and\ \bibinfo {author} {\bibfnamefont {J.}~\bibnamefont
  {Agudo-Canalejo}},\ }\bibfield  {title} {\bibinfo {title} {{Catalysis-Induced
  Phase Separation and Autoregulation of Enzymatic Activity}},\ }\href
  {https://doi.org/10.1103/PHYSREVLETT.129.158101/FIGURES/3/MEDIUM} {\bibfield
  {journal} {\bibinfo  {journal} {Physical Review Letters}\ }\textbf {\bibinfo
  {volume} {129}},\ \bibinfo {pages} {158101} (\bibinfo {year}
  {2022}{\natexlab{a}})}\BibitemShut {NoStop}%
\bibitem [{\citenamefont {Zwicker}(2022)}]{Zwicker2022TheSeparation}%
  \BibitemOpen
  \bibfield  {author} {\bibinfo {author} {\bibfnamefont {D.}~\bibnamefont
  {Zwicker}},\ }\bibfield  {title} {\bibinfo {title} {{The intertwined physics
  of active chemical reactions and phase separation}},\ }\bibfield  {journal}
  {\bibinfo  {journal} {Current Opinion in Colloid and Interface Science}\
  }\textbf {\bibinfo {volume} {61}},\ \href
  {https://doi.org/10.1016/j.cocis.2022.101606} {10.1016/j.cocis.2022.101606}
  (\bibinfo {year} {2022})\BibitemShut {NoStop}%
\bibitem [{\citenamefont {D{\"{o}}rner}\ and\ \citenamefont
  {Hondele}(2024)}]{Dorner2024TheOrganelles}%
  \BibitemOpen
  \bibfield  {author} {\bibinfo {author} {\bibfnamefont {K.}~\bibnamefont
  {D{\"{o}}rner}}\ and\ \bibinfo {author} {\bibfnamefont {M.}~\bibnamefont
  {Hondele}},\ }\bibfield  {title} {\bibinfo {title} {{The Story of RNA
  Unfolded: The Molecular Function of DEAD- and DExH-Box ATPases and Their
  Complex Relationship with Membraneless Organelles}},\ }\bibfield  {journal}
  {\bibinfo  {journal} {Annual Review of Biochemistry}\ }\textbf {\bibinfo
  {volume} {93}},\ \href
  {https://doi.org/10.1146/annurev-biochem-052521-121259}
  {10.1146/annurev-biochem-052521-121259} (\bibinfo {year} {2024})\BibitemShut
  {NoStop}%
\bibitem [{\citenamefont {Zwicker}\ \emph {et~al.}(2016)\citenamefont
  {Zwicker}, \citenamefont {Seyboldt}, \citenamefont {Weber}, \citenamefont
  {Hyman},\ and\ \citenamefont {J{\"{u}}licher}}]{Zwicker2016GrowthProtocells}%
  \BibitemOpen
  \bibfield  {author} {\bibinfo {author} {\bibfnamefont {D.}~\bibnamefont
  {Zwicker}}, \bibinfo {author} {\bibfnamefont {R.}~\bibnamefont {Seyboldt}},
  \bibinfo {author} {\bibfnamefont {C.~A.}\ \bibnamefont {Weber}}, \bibinfo
  {author} {\bibfnamefont {A.~A.}\ \bibnamefont {Hyman}},\ and\ \bibinfo
  {author} {\bibfnamefont {F.}~\bibnamefont {J{\"{u}}licher}},\ }\bibfield
  {title} {\bibinfo {title} {{Growth and division of active droplets provides a
  model for protocells}},\ }\href {https://doi.org/10.1038/nphys3984}
  {\bibfield  {journal} {\bibinfo  {journal} {Nature Physics 2016 13:4}\
  }\textbf {\bibinfo {volume} {13}},\ \bibinfo {pages} {408} (\bibinfo {year}
  {2016})}\BibitemShut {NoStop}%
\bibitem [{\citenamefont {Kirschbaum}\ and\ \citenamefont
  {Zwicker}(2021)}]{Kirschbaum2021ControllingReactions}%
  \BibitemOpen
  \bibfield  {author} {\bibinfo {author} {\bibfnamefont {J.}~\bibnamefont
  {Kirschbaum}}\ and\ \bibinfo {author} {\bibfnamefont {D.}~\bibnamefont
  {Zwicker}},\ }\bibfield  {title} {\bibinfo {title} {{Controlling biomolecular
  condensates via chemical reactions}},\ }\bibfield  {journal} {\bibinfo
  {journal} {Journal of the Royal Society Interface}\ }\textbf {\bibinfo
  {volume} {18}},\ \href {https://doi.org/10.1098/RSIF.2021.0255}
  {10.1098/RSIF.2021.0255} (\bibinfo {year} {2021})\BibitemShut {NoStop}%
\bibitem [{\citenamefont {Bauermann}\ \emph
  {et~al.}(2022{\natexlab{b}})\citenamefont {Bauermann}, \citenamefont
  {Weber},\ and\ \citenamefont {J{\"{u}}licher}}]{Bauermann2022a}%
  \BibitemOpen
  \bibfield  {author} {\bibinfo {author} {\bibfnamefont {J.}~\bibnamefont
  {Bauermann}}, \bibinfo {author} {\bibfnamefont {C.~A.}\ \bibnamefont
  {Weber}},\ and\ \bibinfo {author} {\bibfnamefont {F.}~\bibnamefont
  {J{\"{u}}licher}},\ }\bibfield  {title} {\bibinfo {title} {{Energy and Matter
  Supply for Active Droplets}},\ }\href
  {https://doi.org/10.1002/ANDP.202200132} {\bibfield  {journal} {\bibinfo
  {journal} {Annalen der Physik}\ }\textbf {\bibinfo {volume} {534}},\ \bibinfo
  {pages} {2200132} (\bibinfo {year} {2022}{\natexlab{b}})}\BibitemShut
  {NoStop}%
\bibitem [{\citenamefont {Sastre}\ \emph {et~al.}(2025)\citenamefont {Sastre},
  \citenamefont {Thatte}, \citenamefont {Bergmann}, \citenamefont {Stasi},
  \citenamefont {Tena-Solsona}, \citenamefont {Weber},\ and\ \citenamefont
  {Boekhoven}}]{Sastre2025SizeCells}%
  \BibitemOpen
  \bibfield  {author} {\bibinfo {author} {\bibfnamefont {J.}~\bibnamefont
  {Sastre}}, \bibinfo {author} {\bibfnamefont {A.}~\bibnamefont {Thatte}},
  \bibinfo {author} {\bibfnamefont {A.~M.}\ \bibnamefont {Bergmann}}, \bibinfo
  {author} {\bibfnamefont {M.}~\bibnamefont {Stasi}}, \bibinfo {author}
  {\bibfnamefont {M.}~\bibnamefont {Tena-Solsona}}, \bibinfo {author}
  {\bibfnamefont {C.~A.}\ \bibnamefont {Weber}},\ and\ \bibinfo {author}
  {\bibfnamefont {J.}~\bibnamefont {Boekhoven}},\ }\bibfield  {title} {\bibinfo
  {title} {{Size control and oscillations of active droplets in synthetic
  cells}},\ }\href {https://doi.org/10.1038/s41467-025-57240-8} {\bibfield
  {journal} {\bibinfo  {journal} {Nature Communications 2025 16:1}\ }\textbf
  {\bibinfo {volume} {16}},\ \bibinfo {pages} {1} (\bibinfo {year}
  {2025})}\BibitemShut {NoStop}%
\bibitem [{\citenamefont {Leake}(2013)}]{Leake2013TheTime}%
  \BibitemOpen
  \bibfield  {author} {\bibinfo {author} {\bibfnamefont {M.~C.}\ \bibnamefont
  {Leake}},\ }\bibfield  {title} {\bibinfo {title} {{The physics of life: One
  molecule at a time}},\ }\bibfield  {journal} {\bibinfo  {journal}
  {Philosophical Transactions of the Royal Society B: Biological Sciences}\
  }\textbf {\bibinfo {volume} {368}},\ \href
  {https://doi.org/10.1098/rstb.2012.0248} {10.1098/rstb.2012.0248} (\bibinfo
  {year} {2013})\BibitemShut {NoStop}%
\bibitem [{\citenamefont {Bo}\ \emph {et~al.}(2021)\citenamefont {Bo},
  \citenamefont {Hubatsch}, \citenamefont {Bauermann}, \citenamefont {Weber},\
  and\ \citenamefont {J{\"{u}}licher}}]{Bo2021a}%
  \BibitemOpen
  \bibfield  {author} {\bibinfo {author} {\bibfnamefont {S.}~\bibnamefont
  {Bo}}, \bibinfo {author} {\bibfnamefont {L.}~\bibnamefont {Hubatsch}},
  \bibinfo {author} {\bibfnamefont {J.}~\bibnamefont {Bauermann}}, \bibinfo
  {author} {\bibfnamefont {C.~A.}\ \bibnamefont {Weber}},\ and\ \bibinfo
  {author} {\bibfnamefont {F.}~\bibnamefont {J{\"{u}}licher}},\ }\bibfield
  {title} {\bibinfo {title} {{Stochastic dynamics of single molecules across
  phase boundaries}},\ }\bibfield  {journal} {\bibinfo  {journal} {Physical
  Review Research}\ }\textbf {\bibinfo {volume} {3}},\ \href
  {https://doi.org/10.1103/PhysRevResearch.3.043150}
  {10.1103/PhysRevResearch.3.043150} (\bibinfo {year} {2021})\BibitemShut
  {NoStop}%
\bibitem [{\citenamefont {Heltberg}\ \emph {et~al.}(2021)\citenamefont
  {Heltberg}, \citenamefont {Min{\'{e}}-Hattab}, \citenamefont {Taddei},
  \citenamefont {Walczak},\ and\ \citenamefont
  {Mora}}]{Heltberg2021PhysicalSub-compartments}%
  \BibitemOpen
  \bibfield  {author} {\bibinfo {author} {\bibfnamefont {M.~L.}\ \bibnamefont
  {Heltberg}}, \bibinfo {author} {\bibfnamefont {J.}~\bibnamefont
  {Min{\'{e}}-Hattab}}, \bibinfo {author} {\bibfnamefont {A.}~\bibnamefont
  {Taddei}}, \bibinfo {author} {\bibfnamefont {A.~M.}\ \bibnamefont
  {Walczak}},\ and\ \bibinfo {author} {\bibfnamefont {T.}~\bibnamefont
  {Mora}},\ }\bibfield  {title} {\bibinfo {title} {{Physical observables to
  determine the nature of membrane-less cellular sub-compartments}},\ }\href
  {https://doi.org/10.7554/eLife.69181} {\bibfield  {journal} {\bibinfo
  {journal} {eLife}\ }\textbf {\bibinfo {volume} {10}},\ \bibinfo {pages} {1}
  (\bibinfo {year} {2021})}\BibitemShut {NoStop}%
\bibitem [{\citenamefont {Kent}\ \emph {et~al.}(2020)\citenamefont {Kent},
  \citenamefont {Brown}, \citenamefont {Yang}, \citenamefont {Alsaihati},
  \citenamefont {Tian}, \citenamefont {Wang},\ and\ \citenamefont
  {Ren}}]{Kent}%
  \BibitemOpen
  \bibfield  {author} {\bibinfo {author} {\bibfnamefont {S.}~\bibnamefont
  {Kent}}, \bibinfo {author} {\bibfnamefont {K.}~\bibnamefont {Brown}},
  \bibinfo {author} {\bibfnamefont {C.~h.}\ \bibnamefont {Yang}}, \bibinfo
  {author} {\bibfnamefont {N.}~\bibnamefont {Alsaihati}}, \bibinfo {author}
  {\bibfnamefont {C.}~\bibnamefont {Tian}}, \bibinfo {author} {\bibfnamefont
  {H.}~\bibnamefont {Wang}},\ and\ \bibinfo {author} {\bibfnamefont
  {X.}~\bibnamefont {Ren}},\ }\bibfield  {title} {\bibinfo {title}
  {{Phase-Separated Transcriptional Condensates Accelerate Target-Search
  Process Revealed by Live-Cell Single-Molecule Imaging}},\ }\bibfield
  {journal} {\bibinfo  {journal} {Cell Reports}\ }\textbf {\bibinfo {volume}
  {33}},\ \href {https://doi.org/10.1016/j.celrep.2020.108248}
  {10.1016/j.celrep.2020.108248} (\bibinfo {year} {2020})\BibitemShut {NoStop}%
\bibitem [{\citenamefont {Chappidi}\ \emph {et~al.}(2024)\citenamefont
  {Chappidi}, \citenamefont {Quail}, \citenamefont {Doll}, \citenamefont
  {Vogel}, \citenamefont {Aleksandrov}, \citenamefont {Felekyan}, \citenamefont
  {K{\"{u}}hnemuth}, \citenamefont {Stoynov}, \citenamefont {Seidel},
  \citenamefont {Brugu{\'{e}}s}, \citenamefont {Jahnel}, \citenamefont
  {Franzmann},\ and\ \citenamefont {Alberti}}]{Chappidi2024PARP1-DNAEnds}%
  \BibitemOpen
  \bibfield  {author} {\bibinfo {author} {\bibfnamefont {N.}~\bibnamefont
  {Chappidi}}, \bibinfo {author} {\bibfnamefont {T.}~\bibnamefont {Quail}},
  \bibinfo {author} {\bibfnamefont {S.}~\bibnamefont {Doll}}, \bibinfo {author}
  {\bibfnamefont {L.~T.}\ \bibnamefont {Vogel}}, \bibinfo {author}
  {\bibfnamefont {R.}~\bibnamefont {Aleksandrov}}, \bibinfo {author}
  {\bibfnamefont {S.}~\bibnamefont {Felekyan}}, \bibinfo {author}
  {\bibfnamefont {R.}~\bibnamefont {K{\"{u}}hnemuth}}, \bibinfo {author}
  {\bibfnamefont {S.}~\bibnamefont {Stoynov}}, \bibinfo {author} {\bibfnamefont
  {C.~A.}\ \bibnamefont {Seidel}}, \bibinfo {author} {\bibfnamefont
  {J.}~\bibnamefont {Brugu{\'{e}}s}}, \bibinfo {author} {\bibfnamefont
  {M.}~\bibnamefont {Jahnel}}, \bibinfo {author} {\bibfnamefont {T.~M.}\
  \bibnamefont {Franzmann}},\ and\ \bibinfo {author} {\bibfnamefont
  {S.}~\bibnamefont {Alberti}},\ }\bibfield  {title} {\bibinfo {title}
  {{PARP1-DNA co-condensation drives DNA repair site assembly to prevent
  disjunction of broken DNA ends}},\ }\href
  {https://doi.org/10.1016/j.cell.2024.01.015} {\bibfield  {journal} {\bibinfo
  {journal} {Cell}\ }\textbf {\bibinfo {volume} {187}},\ \bibinfo {pages} {945}
  (\bibinfo {year} {2024})}\BibitemShut {NoStop}%
\bibitem [{\citenamefont {Erkamp}\ \emph {et~al.}(2024)\citenamefont {Erkamp},
  \citenamefont {Farag}, \citenamefont {Qiu}, \citenamefont {Qian},
  \citenamefont {Sneideris}, \citenamefont {Lew}, \citenamefont {Knowles},\
  and\ \citenamefont {Pappu}}]{Erkamp2024DifferentialCondensates}%
  \BibitemOpen
  \bibfield  {author} {\bibinfo {author} {\bibfnamefont {N.~A.}\ \bibnamefont
  {Erkamp}}, \bibinfo {author} {\bibfnamefont {M.}~\bibnamefont {Farag}},
  \bibinfo {author} {\bibfnamefont {Y.}~\bibnamefont {Qiu}}, \bibinfo {author}
  {\bibfnamefont {D.}~\bibnamefont {Qian}}, \bibinfo {author} {\bibfnamefont
  {T.}~\bibnamefont {Sneideris}}, \bibinfo {author} {\bibfnamefont
  {M.}~\bibnamefont {Lew}}, \bibinfo {author} {\bibfnamefont {T.~P.~J.}\
  \bibnamefont {Knowles}},\ and\ \bibinfo {author} {\bibfnamefont {R.~V.}\
  \bibnamefont {Pappu}},\ }\bibfield  {title} {\bibinfo {title} {{Differential
  interactions determine anisotropies at interfaces of RNA-based biomolecular
  condensates}},\ }\bibfield  {journal} {\bibinfo  {journal} {biorXiv}\
  }\textbf {\bibinfo {volume} {2024.08.19}},\ \href
  {https://doi.org/10.1101/2024.08.19.608662} {10.1101/2024.08.19.608662}
  (\bibinfo {year} {2024})\BibitemShut {NoStop}%
\bibitem [{\citenamefont {Min{\'{e}}-Hattab}\ \emph {et~al.}(2021)\citenamefont
  {Min{\'{e}}-Hattab}, \citenamefont {Heltberg}, \citenamefont {Villemeur},
  \citenamefont {Guedj}, \citenamefont {Mora}, \citenamefont {Walczak},
  \citenamefont {Dahan},\ and\ \citenamefont {Taddei}}]{Mine-Hattab2021}%
  \BibitemOpen
  \bibfield  {author} {\bibinfo {author} {\bibfnamefont {J.}~\bibnamefont
  {Min{\'{e}}-Hattab}}, \bibinfo {author} {\bibfnamefont {M.}~\bibnamefont
  {Heltberg}}, \bibinfo {author} {\bibfnamefont {M.}~\bibnamefont {Villemeur}},
  \bibinfo {author} {\bibfnamefont {C.}~\bibnamefont {Guedj}}, \bibinfo
  {author} {\bibfnamefont {T.}~\bibnamefont {Mora}}, \bibinfo {author}
  {\bibfnamefont {A.~M.}\ \bibnamefont {Walczak}}, \bibinfo {author}
  {\bibfnamefont {M.}~\bibnamefont {Dahan}},\ and\ \bibinfo {author}
  {\bibfnamefont {A.}~\bibnamefont {Taddei}},\ }\bibfield  {title} {\bibinfo
  {title} {{Single molecule microscopy reveals key physical features of repair
  foci in living cells.}},\ }\href {https://doi.org/10.7554/eLife.60577}
  {\bibfield  {journal} {\bibinfo  {journal} {eLife}\ }\textbf {\bibinfo
  {volume} {10}},\ \bibinfo {pages} {1} (\bibinfo {year} {2021})}\BibitemShut
  {NoStop}%
\bibitem [{\citenamefont {Lakowicz}(1999)}]{Lakowicz1999}%
  \BibitemOpen
  \bibfield  {author} {\bibinfo {author} {\bibfnamefont {J.}~\bibnamefont
  {Lakowicz}},\ }\href@noop {} {\emph {\bibinfo {title} {{Principles of
  Fluorescence Spectroscopy}}}}\ (\bibinfo  {publisher} {Springer US},\
  \bibinfo {address} {Boston, MA},\ \bibinfo {year} {1999})\BibitemShut
  {NoStop}%
\bibitem [{\citenamefont {Krapivsky}\ \emph {et~al.}(2010)\citenamefont
  {Krapivsky}, \citenamefont {Redner},\ and\ \citenamefont
  {Ben-Naim}}]{Krapivsky2010}%
  \BibitemOpen
  \bibfield  {author} {\bibinfo {author} {\bibfnamefont {P.~L.}\ \bibnamefont
  {Krapivsky}}, \bibinfo {author} {\bibfnamefont {S.}~\bibnamefont {Redner}},\
  and\ \bibinfo {author} {\bibfnamefont {E.}~\bibnamefont {Ben-Naim}},\
  }\href@noop {} {\emph {\bibinfo {title} {{A Kinetic View of Statistical
  Physics}}}}\ (\bibinfo  {publisher} {Cambridge University Press},\ \bibinfo
  {address} {Cambridge},\ \bibinfo {year} {2010})\BibitemShut {NoStop}%
\bibitem [{\citenamefont {De~Gennes}(1979)}]{DeGennes1979ScalingPress.}%
  \BibitemOpen
  \bibfield  {author} {\bibinfo {author} {\bibfnamefont {P.~G.}\ \bibnamefont
  {De~Gennes}},\ }\bibfield  {title} {\bibinfo {title} {{Scaling concepts in
  polymer physics. Cornell university press.}},\ }\href
  {https://www.cornellpress.cornell.edu/book/9780801412035/scaling-concepts-in-polymer-physics/#bookTabs=1}
  {\bibfield  {journal} {\bibinfo  {journal} {Ithaca N.Y.,}\ ,\ \bibinfo
  {pages} {324}} (\bibinfo {year} {1979})}\BibitemShut {NoStop}%
\bibitem [{\citenamefont {Kramer}\ \emph {et~al.}(1984)\citenamefont {Kramer},
  \citenamefont {Green},\ and\ \citenamefont {Palmstr{\o}m}}]{Kramer1984}%
  \BibitemOpen
  \bibfield  {author} {\bibinfo {author} {\bibfnamefont {E.~J.}\ \bibnamefont
  {Kramer}}, \bibinfo {author} {\bibfnamefont {P.}~\bibnamefont {Green}},\ and\
  \bibinfo {author} {\bibfnamefont {C.~J.}\ \bibnamefont {Palmstr{\o}m}},\
  }\bibfield  {title} {\bibinfo {title} {{Interdiffusion and marker movements
  in concentrated polymer-polymer diffusion couples}},\ }\href
  {https://doi.org/10.1016/0032-3861(84)90205-2} {\bibfield  {journal}
  {\bibinfo  {journal} {Polymer}\ }\textbf {\bibinfo {volume} {25}},\ \bibinfo
  {pages} {473} (\bibinfo {year} {1984})}\BibitemShut {NoStop}%
\bibitem [{\citenamefont {Mao}\ \emph {et~al.}(2020)\citenamefont {Mao},
  \citenamefont {Chakraverti-Wuerthwein}, \citenamefont {Gaudio},\ and\
  \citenamefont {Ko{\v{s}}mrlj}}]{Mao2020}%
  \BibitemOpen
  \bibfield  {author} {\bibinfo {author} {\bibfnamefont {S.}~\bibnamefont
  {Mao}}, \bibinfo {author} {\bibfnamefont {M.~S.}\ \bibnamefont
  {Chakraverti-Wuerthwein}}, \bibinfo {author} {\bibfnamefont {H.}~\bibnamefont
  {Gaudio}},\ and\ \bibinfo {author} {\bibfnamefont {A.}~\bibnamefont
  {Ko{\v{s}}mrlj}},\ }\bibfield  {title} {\bibinfo {title} {{Designing the
  Morphology of Separated Phases in Multicomponent Liquid Mixtures}},\
  }\bibfield  {journal} {\bibinfo  {journal} {Physical Review Letters}\
  }\textbf {\bibinfo {volume} {125}},\ \href
  {https://doi.org/10.1103/PhysRevLett.125.218003}
  {10.1103/PhysRevLett.125.218003} (\bibinfo {year} {2020})\BibitemShut
  {NoStop}%
\bibitem [{\citenamefont {Cotton}\ \emph
  {et~al.}(2022{\natexlab{b}})\citenamefont {Cotton}, \citenamefont
  {Golestanian},\ and\ \citenamefont {Agudo-Canalejo}}]{Cotton2022}%
  \BibitemOpen
  \bibfield  {author} {\bibinfo {author} {\bibfnamefont {M.~W.}\ \bibnamefont
  {Cotton}}, \bibinfo {author} {\bibfnamefont {R.}~\bibnamefont
  {Golestanian}},\ and\ \bibinfo {author} {\bibfnamefont {J.}~\bibnamefont
  {Agudo-Canalejo}},\ }\bibfield  {title} {\bibinfo {title} {{Catalysis-Induced
  Phase Separation and Autoregulation of Enzymatic Activity}},\ }\bibfield
  {journal} {\bibinfo  {journal} {Physical Review Letters}\ }\textbf {\bibinfo
  {volume} {129}},\ \href {https://doi.org/10.1103/PhysRevLett.129.158101}
  {10.1103/PhysRevLett.129.158101} (\bibinfo {year}
  {2022}{\natexlab{b}})\BibitemShut {NoStop}%
\bibitem [{\citenamefont {van Kampen}(1984)}]{VanKampen1984}%
  \BibitemOpen
  \bibfield  {author} {\bibinfo {author} {\bibfnamefont {N.}~\bibnamefont {van
  Kampen}},\ }\bibfield  {title} {\bibinfo {title} {{The Gibbs Paradox}},\ }in\
  \href {https://doi.org/10.1016/B978-0-08-026523-0.50020-5} {\emph {\bibinfo
  {booktitle} {Essays in Theoretical Physics}}}\ (\bibinfo  {publisher}
  {Elsevier},\ \bibinfo {year} {1984})\ pp.\ \bibinfo {pages}
  {303--312}\BibitemShut {NoStop}%
\bibitem [{\citenamefont {Gillespie}(1992)}]{gillespie1992rigorous}%
  \BibitemOpen
  \bibfield  {author} {\bibinfo {author} {\bibfnamefont {D.~T.}\ \bibnamefont
  {Gillespie}},\ }\bibfield  {title} {\bibinfo {title} {{A rigorous derivation
  of the chemical master equation}},\ }\href@noop {} {\bibfield  {journal}
  {\bibinfo  {journal} {Physica A: Statistical Mechanics and its Applications}\
  }\textbf {\bibinfo {volume} {188}},\ \bibinfo {pages} {404} (\bibinfo {year}
  {1992})}\BibitemShut {NoStop}%
\bibitem [{\citenamefont {Gardiner}(1985)}]{Gardiner1985}%
  \BibitemOpen
  \bibfield  {author} {\bibinfo {author} {\bibfnamefont {C.}~\bibnamefont
  {Gardiner}},\ }\href@noop {} {\emph {\bibinfo {title} {{Handbook of
  Stochastic Methods for Physics, Chemistry, and the Natural Sciences}}}},\
  \bibinfo {edition} {2nd}\ ed.\ (\bibinfo  {publisher} {Springer},\ \bibinfo
  {address} {Heidelberg},\ \bibinfo {year} {1985})\BibitemShut {NoStop}%
\bibitem [{\citenamefont {Pavliotis}\ and\ \citenamefont
  {Stuart}(2008)}]{pavliotis2008multiscale}%
  \BibitemOpen
  \bibfield  {author} {\bibinfo {author} {\bibfnamefont {G.~A.}\ \bibnamefont
  {Pavliotis}}\ and\ \bibinfo {author} {\bibfnamefont {A.~M.}\ \bibnamefont
  {Stuart}},\ }\href@noop {} {\emph {\bibinfo {title} {{Multiscale methods:
  averaging and homogenization}}}},\ Vol.~\bibinfo {volume} {53}\ (\bibinfo
  {publisher} {Springer},\ \bibinfo {year} {2008})\BibitemShut {NoStop}%
\bibitem [{\citenamefont {Bo}\ and\ \citenamefont {Celani}(2017)}]{Bo2017}%
  \BibitemOpen
  \bibfield  {author} {\bibinfo {author} {\bibfnamefont {S.}~\bibnamefont
  {Bo}}\ and\ \bibinfo {author} {\bibfnamefont {A.}~\bibnamefont {Celani}},\
  }\bibfield  {title} {\bibinfo {title} {{Multiple-scale stochastic processes:
  Decimation, averaging and beyond}},\ }\bibfield  {journal} {\bibinfo
  {journal} {Physics Reports}\ }\textbf {\bibinfo {volume} {670}},\ \href
  {https://doi.org/10.1016/j.physrep.2016.12.003}
  {10.1016/j.physrep.2016.12.003} (\bibinfo {year} {2017})\BibitemShut
  {NoStop}%
\bibitem [{\citenamefont {Aurell}\ and\ \citenamefont {Bo}(2017)}]{Aurell2017}%
  \BibitemOpen
  \bibfield  {author} {\bibinfo {author} {\bibfnamefont {E.}~\bibnamefont
  {Aurell}}\ and\ \bibinfo {author} {\bibfnamefont {S.}~\bibnamefont {Bo}},\
  }\bibfield  {title} {\bibinfo {title} {{Steady diffusion in a drift field: A
  comparison of large-deviation techniques and multiple-scale analysis}},\
  }\bibfield  {journal} {\bibinfo  {journal} {Physical Review E}\ }\textbf
  {\bibinfo {volume} {96}},\ \href {https://doi.org/10.1103/PhysRevE.96.032140}
  {10.1103/PhysRevE.96.032140} (\bibinfo {year} {2017})\BibitemShut {NoStop}%
\bibitem [{\citenamefont {Balzarotti}\ \emph {et~al.}(2017)\citenamefont
  {Balzarotti}, \citenamefont {Eilers}, \citenamefont {Gwosch}, \citenamefont
  {Gynn{\aa}}, \citenamefont {Westphal}, \citenamefont {Stefani}, \citenamefont
  {Elf},\ and\ \citenamefont {Hell}}]{Balzarotti2017NanometerFluxes}%
  \BibitemOpen
  \bibfield  {author} {\bibinfo {author} {\bibfnamefont {F.}~\bibnamefont
  {Balzarotti}}, \bibinfo {author} {\bibfnamefont {Y.}~\bibnamefont {Eilers}},
  \bibinfo {author} {\bibfnamefont {K.~C.}\ \bibnamefont {Gwosch}}, \bibinfo
  {author} {\bibfnamefont {A.~H.}\ \bibnamefont {Gynn{\aa}}}, \bibinfo {author}
  {\bibfnamefont {V.}~\bibnamefont {Westphal}}, \bibinfo {author}
  {\bibfnamefont {F.~D.}\ \bibnamefont {Stefani}}, \bibinfo {author}
  {\bibfnamefont {J.}~\bibnamefont {Elf}},\ and\ \bibinfo {author}
  {\bibfnamefont {S.~W.}\ \bibnamefont {Hell}},\ }\bibfield  {title} {\bibinfo
  {title} {{Nanometer resolution imaging and tracking of fluorescent molecules
  with minimal photon fluxes}},\ }\href
  {https://doi.org/10.1126/SCIENCE.AAK9913/SUPPL{\_}FILE/BALZAROTTI{\_}SM.PDF}
  {\bibfield  {journal} {\bibinfo  {journal} {Science}\ }\textbf {\bibinfo
  {volume} {355}},\ \bibinfo {pages} {606} (\bibinfo {year}
  {2017})}\BibitemShut {NoStop}%
\bibitem [{\citenamefont {Agudo-Canalejo}\ \emph
  {et~al.}(2018{\natexlab{a}})\citenamefont {Agudo-Canalejo}, \citenamefont
  {Illien},\ and\ \citenamefont
  {Golestanian}}]{Agudo-Canalejo2018PhoresisChemotaxis}%
  \BibitemOpen
  \bibfield  {author} {\bibinfo {author} {\bibfnamefont {J.}~\bibnamefont
  {Agudo-Canalejo}}, \bibinfo {author} {\bibfnamefont {P.}~\bibnamefont
  {Illien}},\ and\ \bibinfo {author} {\bibfnamefont {R.}~\bibnamefont
  {Golestanian}},\ }\bibfield  {title} {\bibinfo {title} {{Phoresis and
  Enhanced Diffusion Compete in Enzyme Chemotaxis}},\ }\href
  {https://doi.org/10.1021/ACS.NANOLETT.8B00717/ASSET/IMAGES/LARGE/NL-2018-007178{\_}0002.JPEG}
  {\bibfield  {journal} {\bibinfo  {journal} {Nano Letters}\ }\textbf {\bibinfo
  {volume} {18}},\ \bibinfo {pages} {2711} (\bibinfo {year}
  {2018}{\natexlab{a}})}\BibitemShut {NoStop}%
\bibitem [{\citenamefont {Agudo-Canalejo}\ \emph
  {et~al.}(2018{\natexlab{b}})\citenamefont {Agudo-Canalejo}, \citenamefont
  {Adeleke-Larodo}, \citenamefont {Illien},\ and\ \citenamefont
  {Golestanian}}]{Agudo-Canalejo2018EnhancedNanoscale}%
  \BibitemOpen
  \bibfield  {author} {\bibinfo {author} {\bibfnamefont {J.}~\bibnamefont
  {Agudo-Canalejo}}, \bibinfo {author} {\bibfnamefont {T.}~\bibnamefont
  {Adeleke-Larodo}}, \bibinfo {author} {\bibfnamefont {P.}~\bibnamefont
  {Illien}},\ and\ \bibinfo {author} {\bibfnamefont {R.}~\bibnamefont
  {Golestanian}},\ }\bibfield  {title} {\bibinfo {title} {{Enhanced Diffusion
  and Chemotaxis at the Nanoscale}},\ }\href
  {https://doi.org/10.1021/ACS.ACCOUNTS.8B00280/ASSET/IMAGES/MEDIUM/AR-2018-00280H{\_}0007.GIF}
  {\bibfield  {journal} {\bibinfo  {journal} {Accounts of Chemical Research}\
  }\textbf {\bibinfo {volume} {51}},\ \bibinfo {pages} {2365} (\bibinfo {year}
  {2018}{\natexlab{b}})}\BibitemShut {NoStop}%
\bibitem [{\citenamefont {Agudo-Canalejo}\ \emph {et~al.}(2020)\citenamefont
  {Agudo-Canalejo}, \citenamefont {Illien},\ and\ \citenamefont
  {Golestanian}}]{Agudo-Canalejo2020CooperativelyProteins}%
  \BibitemOpen
  \bibfield  {author} {\bibinfo {author} {\bibfnamefont {J.}~\bibnamefont
  {Agudo-Canalejo}}, \bibinfo {author} {\bibfnamefont {P.}~\bibnamefont
  {Illien}},\ and\ \bibinfo {author} {\bibfnamefont {R.}~\bibnamefont
  {Golestanian}},\ }\bibfield  {title} {\bibinfo {title} {{Cooperatively
  enhanced reactivity and “stabilitaxis” of dissociating oligomeric
  proteins}},\ }\href
  {https://doi.org/10.1073/PNAS.1919635117/SUPPL{\_}FILE/PNAS.1919635117.SAPP.PDF}
  {\bibfield  {journal} {\bibinfo  {journal} {Proceedings of the National
  Academy of Sciences of the United States of America}\ }\textbf {\bibinfo
  {volume} {117}},\ \bibinfo {pages} {11894} (\bibinfo {year}
  {2020})}\BibitemShut {NoStop}%
\bibitem [{\citenamefont {Feng}\ and\ \citenamefont
  {Gilson}(2020)}]{Feng2020EnhancedEnzymes}%
  \BibitemOpen
  \bibfield  {author} {\bibinfo {author} {\bibfnamefont {M.}~\bibnamefont
  {Feng}}\ and\ \bibinfo {author} {\bibfnamefont {M.~K.}\ \bibnamefont
  {Gilson}},\ }\bibfield  {title} {\bibinfo {title} {{Enhanced Diffusion and
  Chemotaxis of Enzymes}},\ }\href
  {https://doi.org/10.1146/ANNUREV-BIOPHYS-121219-081535/CITE/REFWORKS}
  {\bibfield  {journal} {\bibinfo  {journal} {Annual Review of Biophysics}\
  }\textbf {\bibinfo {volume} {49}},\ \bibinfo {pages} {87} (\bibinfo {year}
  {2020})}\BibitemShut {NoStop}%
\bibitem [{\citenamefont {Taylor}(1953)}]{Taylor1953}%
  \BibitemOpen
  \bibfield  {author} {\bibinfo {author} {\bibfnamefont {G.~I.}\ \bibnamefont
  {Taylor}},\ }\bibfield  {title} {\bibinfo {title} {{Dispersion of soluble
  matter in solvent flowing slowly through a tube}},\ }\href
  {https://doi.org/10.1098/rspa.1953.0139} {\bibfield  {journal} {\bibinfo
  {journal} {Proceedings of the Royal Society of London. Series A. Mathematical
  and Physical Sciences}\ }\textbf {\bibinfo {volume} {219}},\ \bibinfo {pages}
  {186} (\bibinfo {year} {1953})}\BibitemShut {NoStop}%
\bibitem [{\citenamefont {Mysels}(1956)}]{Mysels1956}%
  \BibitemOpen
  \bibfield  {author} {\bibinfo {author} {\bibfnamefont {K.~J.}\ \bibnamefont
  {Mysels}},\ }\bibfield  {title} {\bibinfo {title} {{Electrodiffusion: A
  fluctuation method for measuring reaction rates}},\ }\href
  {https://doi.org/10.1063/1.1742480} {\bibfield  {journal} {\bibinfo
  {journal} {The Journal of Chemical Physics}\ }\textbf {\bibinfo {volume}
  {24}},\ \bibinfo {pages} {371} (\bibinfo {year} {1956})}\BibitemShut
  {NoStop}%
\bibitem [{\citenamefont {Aurell}\ \emph {et~al.}(2016)\citenamefont {Aurell},
  \citenamefont {Bo}, \citenamefont {Dias}, \citenamefont {Eichhorn},\ and\
  \citenamefont {Marino}}]{Aurell2016a}%
  \BibitemOpen
  \bibfield  {author} {\bibinfo {author} {\bibfnamefont {E.}~\bibnamefont
  {Aurell}}, \bibinfo {author} {\bibfnamefont {S.}~\bibnamefont {Bo}}, \bibinfo
  {author} {\bibfnamefont {M.}~\bibnamefont {Dias}}, \bibinfo {author}
  {\bibfnamefont {R.}~\bibnamefont {Eichhorn}},\ and\ \bibinfo {author}
  {\bibfnamefont {R.}~\bibnamefont {Marino}},\ }\bibfield  {title} {\bibinfo
  {title} {{Diffusion of a Brownian ellipsoid in a force field}},\ }\bibfield
  {journal} {\bibinfo  {journal} {EPL}\ }\textbf {\bibinfo {volume} {114}},\
  \href {https://doi.org/10.1209/0295-5075/114/30005}
  {10.1209/0295-5075/114/30005} (\bibinfo {year} {2016})\BibitemShut {NoStop}%
\bibitem [{\citenamefont {Kahlen}\ \emph {et~al.}(2017)\citenamefont {Kahlen},
  \citenamefont {Engel},\ and\ \citenamefont {Van Den~Broeck}}]{Kahlen2017}%
  \BibitemOpen
  \bibfield  {author} {\bibinfo {author} {\bibfnamefont {M.}~\bibnamefont
  {Kahlen}}, \bibinfo {author} {\bibfnamefont {A.}~\bibnamefont {Engel}},\ and\
  \bibinfo {author} {\bibfnamefont {C.}~\bibnamefont {Van Den~Broeck}},\
  }\bibfield  {title} {\bibinfo {title} {{Large deviations in Taylor
  dispersion}},\ }\href {https://doi.org/10.1103/PhysRevE.95.012144} {\bibfield
   {journal} {\bibinfo  {journal} {Physical Review E}\ }\textbf {\bibinfo
  {volume} {95}},\ \bibinfo {pages} {1} (\bibinfo {year} {2017})}\BibitemShut
  {NoStop}%
\bibitem [{\citenamefont {Pietzonka}\ \emph {et~al.}(2016)\citenamefont
  {Pietzonka}, \citenamefont {Kleinbeck},\ and\ \citenamefont
  {Seifert}}]{Pietzonka2016}%
  \BibitemOpen
  \bibfield  {author} {\bibinfo {author} {\bibfnamefont {P.}~\bibnamefont
  {Pietzonka}}, \bibinfo {author} {\bibfnamefont {K.}~\bibnamefont
  {Kleinbeck}},\ and\ \bibinfo {author} {\bibfnamefont {U.}~\bibnamefont
  {Seifert}},\ }\bibfield  {title} {\bibinfo {title} {{Extreme fluctuations of
  active Brownian motion}},\ }\href
  {https://doi.org/10.1088/1367-2630/18/5/052001} {\bibfield  {journal}
  {\bibinfo  {journal} {New Journal of Physics}\ }\textbf {\bibinfo {volume}
  {18}},\ \bibinfo {pages} {1} (\bibinfo {year} {2016})}\BibitemShut {NoStop}%
\bibitem [{\citenamefont {Agudo-Canalejo}\ and\ \citenamefont
  {Golestanian}(2020)}]{Agudo-Canalejo2020DiffusionEnzymes}%
  \BibitemOpen
  \bibfield  {author} {\bibinfo {author} {\bibfnamefont {J.}~\bibnamefont
  {Agudo-Canalejo}}\ and\ \bibinfo {author} {\bibfnamefont {R.}~\bibnamefont
  {Golestanian}},\ }\bibfield  {title} {\bibinfo {title} {{Diffusion and steady
  state distributions of flexible chemotactic enzymes}},\ }\href
  {https://doi.org/10.1140/epjst/e2020-900224-3} {\bibfield  {journal}
  {\bibinfo  {journal} {European Physical Journal: Special Topics}\ }\textbf
  {\bibinfo {volume} {229}},\ \bibinfo {pages} {2791} (\bibinfo {year}
  {2020})}\BibitemShut {NoStop}%
\bibitem [{\citenamefont {Vilquin}\ \emph {et~al.}(2023)\citenamefont
  {Vilquin}, \citenamefont {Bertin}, \citenamefont {Rapha{\"{e}}l},
  \citenamefont {Dean}, \citenamefont {Salez},\ and\ \citenamefont
  {McGraw}}]{Vilquin2023NanoparticleBoundary}%
  \BibitemOpen
  \bibfield  {author} {\bibinfo {author} {\bibfnamefont {A.}~\bibnamefont
  {Vilquin}}, \bibinfo {author} {\bibfnamefont {V.}~\bibnamefont {Bertin}},
  \bibinfo {author} {\bibfnamefont {E.}~\bibnamefont {Rapha{\"{e}}l}}, \bibinfo
  {author} {\bibfnamefont {D.~S.}\ \bibnamefont {Dean}}, \bibinfo {author}
  {\bibfnamefont {T.}~\bibnamefont {Salez}},\ and\ \bibinfo {author}
  {\bibfnamefont {J.~D.}\ \bibnamefont {McGraw}},\ }\bibfield  {title}
  {\bibinfo {title} {{Nanoparticle Taylor Dispersion Near Charged Surfaces with
  an Open Boundary}},\ }\href {https://doi.org/10.1103/PhysRevLett.130.038201}
  {\bibfield  {journal} {\bibinfo  {journal} {Physical Review Letters}\
  }\textbf {\bibinfo {volume} {130}},\ \bibinfo {pages} {038201} (\bibinfo
  {year} {2023})}\BibitemShut {NoStop}%
\bibitem [{\citenamefont {Klosin}\ \emph {et~al.}(2020)\citenamefont {Klosin},
  \citenamefont {Oltsch}, \citenamefont {Harmon}, \citenamefont {Honigmann},
  \citenamefont {J{\"{u}}licher}, \citenamefont {Hyman},\ and\ \citenamefont
  {Zechner}}]{Klosin2020}%
  \BibitemOpen
  \bibfield  {author} {\bibinfo {author} {\bibfnamefont {A.}~\bibnamefont
  {Klosin}}, \bibinfo {author} {\bibfnamefont {F.}~\bibnamefont {Oltsch}},
  \bibinfo {author} {\bibfnamefont {T.}~\bibnamefont {Harmon}}, \bibinfo
  {author} {\bibfnamefont {A.}~\bibnamefont {Honigmann}}, \bibinfo {author}
  {\bibfnamefont {F.}~\bibnamefont {J{\"{u}}licher}}, \bibinfo {author}
  {\bibfnamefont {A.~A.}\ \bibnamefont {Hyman}},\ and\ \bibinfo {author}
  {\bibfnamefont {C.}~\bibnamefont {Zechner}},\ }\bibfield  {title} {\bibinfo
  {title} {{Phase separation provides a mechanism to reduce noise in cells}},\
  }\href {https://doi.org/10.1126/science.aav6691} {\bibfield  {journal}
  {\bibinfo  {journal} {Science}\ }\textbf {\bibinfo {volume} {367}},\ \bibinfo
  {pages} {464} (\bibinfo {year} {2020})}\BibitemShut {NoStop}%
\bibitem [{\citenamefont {Deviri}\ and\ \citenamefont
  {Safran}(2021)}]{Deviri2021}%
  \BibitemOpen
  \bibfield  {author} {\bibinfo {author} {\bibfnamefont {D.}~\bibnamefont
  {Deviri}}\ and\ \bibinfo {author} {\bibfnamefont {S.~A.}\ \bibnamefont
  {Safran}},\ }\bibfield  {title} {\bibinfo {title} {{Physical theory of
  biological noise buffering by multicomponent phase separation}},\ }\bibfield
  {journal} {\bibinfo  {journal} {Proceedings of the National Academy of
  Sciences of the United States of America}\ }\textbf {\bibinfo {volume}
  {118}},\ \href {https://doi.org/10.1073/pnas.2100099118}
  {10.1073/pnas.2100099118} (\bibinfo {year} {2021})\BibitemShut {NoStop}%
\bibitem [{\citenamefont {Zechner}\ and\ \citenamefont
  {J{\"{u}}licher}(2025)}]{Zechner2025ConcentrationSystems}%
  \BibitemOpen
  \bibfield  {author} {\bibinfo {author} {\bibfnamefont {C.}~\bibnamefont
  {Zechner}}\ and\ \bibinfo {author} {\bibfnamefont {F.}~\bibnamefont
  {J{\"{u}}licher}},\ }\bibfield  {title} {\bibinfo {title} {{Concentration
  buffering and noise reduction in non-equilibrium phase-separating systems}},\
  }\href {https://doi.org/10.1016/j.cels.2025.101168} {\bibfield  {journal}
  {\bibinfo  {journal} {Cell Systems}\ }\textbf {\bibinfo {volume} {16}},\
  \bibinfo {pages} {101168} (\bibinfo {year} {2025})}\BibitemShut {NoStop}%
\bibitem [{\citenamefont {Zhang}\ \emph {et~al.}(2024)\citenamefont {Zhang},
  \citenamefont {Pyo}, \citenamefont {Kliegman}, \citenamefont {Jiang},
  \citenamefont {Brangwynne}, \citenamefont {Stone},\ and\ \citenamefont
  {Wingreen}}]{Zhang2024TheCondensatesb}%
  \BibitemOpen
  \bibfield  {author} {\bibinfo {author} {\bibfnamefont {Y.}~\bibnamefont
  {Zhang}}, \bibinfo {author} {\bibfnamefont {A.~G.}\ \bibnamefont {Pyo}},
  \bibinfo {author} {\bibfnamefont {R.}~\bibnamefont {Kliegman}}, \bibinfo
  {author} {\bibfnamefont {Y.}~\bibnamefont {Jiang}}, \bibinfo {author}
  {\bibfnamefont {C.~P.}\ \bibnamefont {Brangwynne}}, \bibinfo {author}
  {\bibfnamefont {H.~A.}\ \bibnamefont {Stone}},\ and\ \bibinfo {author}
  {\bibfnamefont {N.~S.}\ \bibnamefont {Wingreen}},\ }\bibfield  {title}
  {\bibinfo {title} {{The exchange dynamics of biomolecular condensates}},\
  }\bibfield  {journal} {\bibinfo  {journal} {eLife}\ }\textbf {\bibinfo
  {volume} {12}},\ \href {https://doi.org/10.7554/ELIFE.91680}
  {10.7554/ELIFE.91680} (\bibinfo {year} {2024})\BibitemShut {NoStop}%
\bibitem [{\citenamefont {Hubatsch}\ \emph {et~al.}(2024)\citenamefont
  {Hubatsch}, \citenamefont {Bo}, \citenamefont {Harmon}, \citenamefont
  {Hyman}, \citenamefont {Weber},\ and\ \citenamefont
  {J{\"{u}}licher}}]{Hubatsch2024TransportPhases}%
  \BibitemOpen
  \bibfield  {author} {\bibinfo {author} {\bibfnamefont {L.}~\bibnamefont
  {Hubatsch}}, \bibinfo {author} {\bibfnamefont {S.}~\bibnamefont {Bo}},
  \bibinfo {author} {\bibfnamefont {T.~S.}\ \bibnamefont {Harmon}}, \bibinfo
  {author} {\bibfnamefont {A.~A.}\ \bibnamefont {Hyman}}, \bibinfo {author}
  {\bibfnamefont {C.~A.}\ \bibnamefont {Weber}},\ and\ \bibinfo {author}
  {\bibfnamefont {F.}~\bibnamefont {J{\"{u}}licher}},\ }\bibfield  {title}
  {\bibinfo {title} {{Transport kinetics across interfaces between coexisting
  liquid phases}},\ }\bibfield  {journal} {\bibinfo  {journal} {bioRxiv}\
  }\href {https://doi.org/10.1101/2024.12.20.629798}
  {10.1101/2024.12.20.629798} (\bibinfo {year} {2024})\BibitemShut {NoStop}%
\bibitem [{\citenamefont
  {Hubatsch}(2025)}]{Hubatsch2025Https://git.mpi-cbg.de/hubatsch/chemical_reactions}%
  \BibitemOpen
  \bibfield  {author} {\bibinfo {author} {\bibfnamefont {L.}~\bibnamefont
  {Hubatsch}},\ }\href {https://git.mpi-cbg.de/hubatsch/chemical_reactions}
  {\bibinfo {title} {{https://git.mpi-cbg.de/hubatsch/chemical{\_}reactions}}}
  (\bibinfo {year} {2025})\BibitemShut {NoStop}%
\bibitem [{\citenamefont {Kloeden}\ and\ \citenamefont
  {Platen}(1992)}]{Kloeden1992}%
  \BibitemOpen
  \bibfield  {author} {\bibinfo {author} {\bibfnamefont {P.~E.}\ \bibnamefont
  {Kloeden}}\ and\ \bibinfo {author} {\bibfnamefont {E.}~\bibnamefont
  {Platen}},\ }\href {https://doi.org/10.1007/978-3-662-12616-5} {\emph
  {\bibinfo {title} {Numerical Solution of Stochastic Differential
  Equations}}},\ \bibinfo {edition} {1st}\ ed.\ (\bibinfo  {publisher}
  {Springer-Verlag, Berlin--Heidelberg},\ \bibinfo {address} {Berlin
  Heidelberg},\ \bibinfo {year} {1992})\BibitemShut {NoStop}%
\bibitem [{\citenamefont {Ragwitz}\ and\ \citenamefont
  {Kantz}(2001)}]{Ragwitz2001IndispensableData}%
  \BibitemOpen
  \bibfield  {author} {\bibinfo {author} {\bibfnamefont {M.}~\bibnamefont
  {Ragwitz}}\ and\ \bibinfo {author} {\bibfnamefont {H.}~\bibnamefont
  {Kantz}},\ }\bibfield  {title} {\bibinfo {title} {{Indispensable Finite Time
  Corrections for Fokker-Planck Equations from Time Series Data}},\ }\href
  {https://doi.org/10.1103/PhysRevLett.87.254501} {\bibfield  {journal}
  {\bibinfo  {journal} {Physical Review Letters}\ }\textbf {\bibinfo {volume}
  {87}},\ \bibinfo {pages} {254501} (\bibinfo {year} {2001})}\BibitemShut
  {NoStop}%
\end{thebibliography}%

\end{document}